\documentclass[preprint,12pt]{elsarticle}

\usepackage{amssymb}
\usepackage{amsmath}
\usepackage{booktabs}
\usepackage{array}
\usepackage{ragged2e}

\journal{Computers in Biology and Medicine}

\begin{document}

\begin{frontmatter}



\title{Deep Neural Network Architectures for Electrocardiogram Classification: A Comprehensive Evaluation} 




\author[label1]{Yun Song\corref{cor1}\fnref{fn1}}
\ead{ysong96@fordham.edu} 

\author[label2]{Wenjia Zheng\corref{cor1}\fnref{fn1}}
\ead{wzheng33@fordham.edu} 

\author[label3]{Tiedan Chen}
\author[label4]{Ziyu Wang}
\author[label5]{Jiazhao Shi}
\author[label6]{Yisong Chen}

\cortext[cor1]{Corresponding author}
\fntext[fn1]{These authors contributed equally to this work.}

\affiliation[label1]{organization={Independent Researcher},
            city={Shanghai},
            postcode={200000}, 
            country={China}}

\affiliation[label2]{organization={Independent Researcher},
            city={Brooklyn},
            state={NY},
            postcode={11217},
            country={USA}}

\affiliation[label3]{organization={Department of Materials Engineering, Xi'an Polytechnic University},
            addressline={No. 19 Jinhua South Road},
            city={Xi'an},
            postcode={710048},
            country={China}}

\affiliation[label4]{organization={Independent Researcher},
            city={Washington D.C.},
            postcode={20001},
            country={USA}}

\affiliation[label5]{organization={Tandon School of Engineering, New York University},
            addressline={6 Metrotech Center},
            city={Brooklyn},
            state={NY},
            postcode={11201},
            country={USA}}

\affiliation[label6]{organization={College of Computing, Georgia Institute of Technology},
            addressline={801 Atlantic Dr NW},
            city={Atlanta},
            state={GA},
            postcode={30332},
            country={USA}}

\begin{abstract}
With the rising prevalence of cardiovascular diseases, electrocardiograms (ECG) remain essential for the non-invasive detection of cardiac abnormalities. This study presents a comprehensive evaluation of deep neural network architectures for automated arrhythmia classification, integrating temporal modeling, attention mechanisms, and ensemble strategies. To address data scarcity in minority classes, the MIT-BIH Arrhythmia dataset was augmented using a Generative Adversarial Network (GAN). We developed and compared four distinct architectures, including Convolutional Neural Networks (CNN), CNN combined with Long Short-Term Memory (CNN-LSTM), CNN-LSTM with Attention, and 1D Residual Networks (ResNet-1D), to capture both local morphological features and long-term temporal dependencies. Performance was rigorously evaluated using accuracy, F1-score, and Area Under the Curve (AUC) with 95\% confidence intervals to ensure statistical robustness, while Gradient-weighted Class Activation Mapping (Grad-CAM) was employed to validate model interpretability. Experimental results indicate that the CNN-LSTM model achieved the optimal stand-alone balance between sensitivity and specificity, yielding an F1-score of 0.951. Conversely, the CNN-LSTM-Attention and ResNet-1D models exhibited higher sensitivity to class imbalance. To mitigate this, a dynamic ensemble fusion strategy was introduced; specifically, the \textit{Top2-Weighted} ensemble achieved the highest overall performance with an F1-score of 0.958. These findings demonstrate that leveraging complementary deep architectures significantly enhances classification reliability, providing a robust and interpretable foundation for intelligent arrhythmia detection systems.
\end{abstract}

\begin{graphicalabstract}
    \centering
    \includegraphics[width=\textwidth]{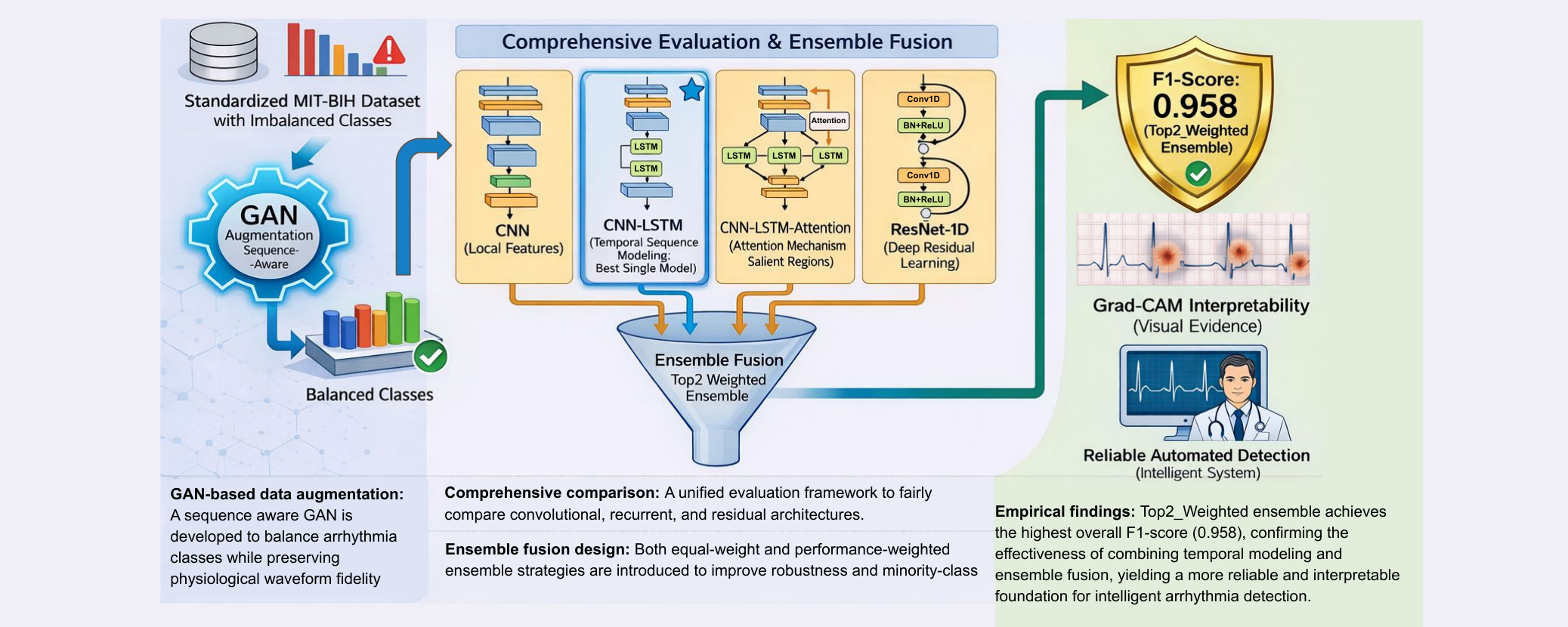}
\end{graphicalabstract}

\begin{highlights}
\item Evaluated four deep neural architectures for automated ECG classification
\item GAN-based data augmentation mitigates class imbalance for minority arrhythmias
\item CNN-LSTM achieves the optimal balance between sensitivity and specificity
\item Attention and ResNet models exhibit higher variability on minority classes
\item \textit{Top2-Weighted} ensemble fusion yields the highest overall F1-score of 0.958
\end{highlights}


\end{frontmatter}



\section{Introduction}

Cardiovascular diseases (CVDs) remain the leading cause of death worldwide, 
accounting for an estimated 19.8 million deaths in 2022, or roughly 32\% of all 
global deaths~\cite{who2025}. The majority of these fatalities are caused by heart 
attacks and strokes, particularly in low- and middle-income countries where early 
diagnosis and preventive care remain limited. As most CVDs are preventable 
through lifestyle and risk-factor management, early detection through 
electrocardiogram (ECG) monitoring plays a crucial role in reducing morbidity 
and mortality.

The ECG is one of the most effective and widely used non-invasive diagnostic 
tools for detecting cardiac abnormalities such as arrhythmias, myocardial 
ischemia, and conduction disorders. By recording the heart's electrical activity, 
it provides a dynamic reflection of cardiac rhythm and morphology, enabling 
timely and accurate diagnosis. However, manual ECG interpretation is 
labor-intensive, prone to inter-observer variability, and increasingly 
impractical for large-scale or long-term monitoring such as 24-hour Holter 
recordings.

These limitations highlight the urgent need for automated, accurate, and reliable 
ECG classification systems that can assist clinicians in real-time diagnosis, 
reduce diagnostic variability, and enable large-scale screening and remote 
health monitoring.

Recent advances in deep learning have revolutionized automated ECG analysis by 
enabling end-to-end learning of morphological and temporal signal 
representations. Convolutional Neural Networks (CNNs) effectively capture local 
waveform features, while Recurrent Neural Networks (RNNs), particularly Long 
Short-Term Memory (LSTM) networks, model sequential dependencies across 
heartbeats. More recently, attention mechanisms have been introduced to 
dynamically emphasize diagnostically salient temporal regions, improving 
interpretability and sensitivity to subtle abnormalities. Despite these 
successes, existing studies often evaluate models under inconsistent settings 
and rarely investigate how architectural design choices affect robustness and 
class-wise balance, especially under severe data imbalance.

To address these challenges, this study conducts a comprehensive evaluation of 
four representative deep learning architectures for ECG classification, 
including CNN, CNN-LSTM, CNN-LSTM-Attention, and ResNet-1D on the standardized 
MIT-BIH Arrhythmia dataset. A Generative Adversarial Network (GAN) is employed to synthesize minority-class samples and mitigate imbalance. The attention 
mechanism is incorporated as one architectural variant to evaluate its 
effectiveness in highlighting diagnostically relevant regions of the ECG signal. 
Based on the results of individual models, several ensemble strategies are 
further designed to enhance model robustness, class-wise balance, and 
generalization performance. Both equal-weighted and performance-weighted fusion 
approaches are explored, aiming to leverage complementary strengths across 
architectures and improve overall classification stability.

The key contributions of this work are summarized as follows:

\begin{enumerate}
    \item \textbf{GAN-based data augmentation:} A sequence-aware GAN is adopted 
    to balance arrhythmia classes while preserving physiological waveform 
    fidelity.
    \item \textbf{Comprehensive comparison:} A unified evaluation framework is 
    established to fairly assess architectural differences under consistent 
    data-processing and training conditions.
    \item \textbf{Ensemble fusion design:} Both equal-weight and performance-weighted ensemble strategies are introduced to improve robustness and minority-class sensitivity.
    \item \textbf{Empirical findings:} Experimental results show that the CNN-LSTM network provides the most balanced and stable single-model performance, while the \textit{Top2-Weighted} ensemble achieves the highest overall F1-score (0.958), confirming the effectiveness of combining temporal modeling and ensemble fusion.  
\end{enumerate}

Overall, this study provides a systematic benchmark of deep neural architectures for ECG classification and demonstrates that integrating temporal sequence modeling, attention mechanisms, and ensemble fusion yields a more reliable and interpretable foundation for intelligent arrhythmia detection.

\section{Related Work}

ECG signals, generated by the depolarization and repolarization cycles of myocardial cells, provide vital information for assessing cardiac function \cite{kligfield2007ecg}. Automated ECG classification research began in the 1970s with rule-based and template-matching methods for QRS detection \cite{pan1985}. In 1975, the Massachusetts Institute of Technology (MIT) and Beth Israel Hospital jointly established the MIT-BIH Arrhythmia Database \cite{moody2001}, which later became a foundational benchmark for automated ECG analysis \cite{goldberger2000}.

In earlier studies, ECG classification primarily relied on handcrafted feature engineering. Commonly used time-domain features include RR intervals, QRS duration, and heart rate variability (HRV) \cite{pan1985}. In the frequency domain, researchers utilized power spectral density and Fourier transform coefficients, while wavelet transform was widely applied to capture the multiscale energy distribution of P waves, QRS complexes, and T waves \cite{addison2005}. Nonlinear dynamics–based features such as sample entropy, approximate entropy, fractal dimension, and Lempel–Ziv complexity were also explored to describe the complexity and irregularity of ECG signals \cite{richman2000}. Based on these manually extracted features, various machine learning classifiers were employed, including Support Vector Machines (SVM), k-Nearest Neighbors (KNN), decision trees, random forests \cite{cortes1995,breiman2001}, and ensemble methods such as AdaBoost and XGBoost \cite{chen2016xgboost}. Although these approaches offered advantages such as interpretability, short training time, and effectiveness in small-sample settings, their performance was limited by reliance on handcrafted features, restricted generalization ability, and insufficient robustness when dealing with complex ECG morphologies.

With the emergence of deep learning, end-to-end architectures increasingly replaced traditional feature engineering. Acharya et al. \cite{acharya2017a} demonstrated that convolutional neural networks can automatically learn discriminative features from tachycardia ECG segments, enabling effective end-to-end arrhythmia detection. In 2019, Hannun et al. \cite{hannun2019deep} introduced a CNN-based arrhythmia classifier that achieved cardiologist-level performance on the MIT-BIH dataset, marking a milestone for automated ECG interpretation. Similarly, Attia et al. \cite{attia2019} demonstrated that deep neural networks could infer hidden patient attributes such as age and sex from ECGs. CNNs excel at capturing spatial morphology, whereas recurrent networks such as LSTMs model temporal dependencies across consecutive beats. Despite these advances, most existing architectures treat all time steps equally, without mechanisms to emphasize diagnostically salient waveform regions.

To address the limitations of conventional deep learning models that treat all time steps or leads uniformly, recent studies have incorporated attention mechanisms into ECG classification. Multi-lead ECG classification methods have been enhanced with channel-wise attention modules to adaptively learn the relative importance of different leads and improve discriminative feature extraction from multi-lead signals \cite{tung2022multi}. Hierarchical attention mechanisms have also been proposed to capture both local and contextual dependencies in ECG sequences, enabling more effective fusion of temporal and structural information for arrhythmia recognition \cite{islam2023hardc}. In addition, graph-based approaches have been leveraged to model complex relationships among ECG signal components by representing ECG features as graph structures and applying graph neural networks with attention operators, demonstrating improved robustness in arrhythmia classification tasks \cite{strodthoff2020ptbxl}. These advancements illustrate that attention mechanisms not only enhance classification performance but also support more interpretable decision processes by highlighting diagnostically salient waveform segments across leads.

Beyond algorithmic progress, bridging the gap between high-performance deep learning models and practical clinical deployment requires robust system-level support. While deep neural networks offer superior diagnostic accuracy, their intensive computational demands often impede real-time performance in resource-constrained environments. To address this, recent research has focused on optimizing the underlying infrastructure. For instance, target-based \cite{zheng2019resource} and elastic resource configuration frameworks \cite{zheng2019flowcon} have been proposed to dynamically redistribute computational resources, accelerating training convergence without sacrificing accuracy. In terms of inference deployment, Differentiated Quality-of-Experience (QoE) scheduling and speculative container management algorithms have been developed to manage concurrent tasks under strict latency constraints in cloud clusters \cite{mao2023cloud, mao2022speculative, mao2023elastic}. Moreover, task-specific architectural principles have been explored in related sequential domains, integrating diverse feature representations to achieve state-of-the-art performance \cite{yuan2021parallel}. Collectively, these system-level and architectural contributions provide the necessary foundation for deploying complex ECG classification models into scalable, real-world medical systems.

Despite substantial progress, most prior work focuses on developing individual architectures rather than systematically comparing them under consistent experimental conditions. Few studies analyze the relative effectiveness of CNN, CNN-LSTM, CNN-LSTM-Attention, and ResNet-1D using the same dataset, nor do they investigate ensemble integration to enhance class balance and robustness. This study aims to bridge this gap through a comprehensive comparative analysis and ensemble-based optimization framework for automated ECG classification.

\section{Methodology}
\subsection{Dataset Preprocessing for ECG Analysis}

This section describes the dataset, preprocessing pipeline, and normalization 
steps applied prior to model training.

\subsection{Dataset and Preprocessing}
\label{sec:dataset_intro}

This study utilizes the MIT-BIH Arrhythmia Database~\cite{moody2001}, a standard benchmark for ECG signal analysis accessible via PhysioNet~\cite{goldberger2000}. The following sections detail the dataset specifications, the category selection protocol, and the preprocessing pipeline designed to optimize model training.

\subsubsection{Dataset Overview}
The database comprises 48 half-hour excerpts of two-channel ambulatory ECG recordings, obtained from 47 distinct subjects. The signals were digitized at a sampling rate of 360 Hz per channel with 11-bit resolution over a 10 mV range. This temporal resolution is sufficient to capture fine-grained morphological details of key waveforms, including the P wave, QRS complex, and T wave. Most records utilize the Modified Limb Lead II (MLII), which highlights the QRS complex, accompanied by a precordial lead (typically V1). Consistent with standard literature practices, this study primarily utilizes the MLII lead for feature extraction and classification.

\subsubsection{Category Selection and Label Mapping}
\label{sec:label_mapping}

The MIT-BIH dataset contains detailed annotations for over a dozen distinct heartbeat types. Instead of aggregating these into broad superclasses, this study focuses on five specific, clinically significant arrhythmia morphologies. To align with the model's training objectives, we extracted and mapped the heartbeats as follows:

\begin{itemize}
    \item \textbf{Normal (N):} Comprises Normal beats (\texttt{N}), as well as Left (\texttt{L}) and Right (\texttt{R}) Bundle Branch Block beats.
    
    \item \textbf{Atrial Premature (A):} Corresponds strictly to the Atrial Premature beats (annotated as \texttt{A}).
    
    \item \textbf{Premature Ventricular Contraction (V):} Corresponds to the Premature Ventricular Contraction beats (annotated as \texttt{V}).
    
    \item \textbf{Fusion of Paced and Normal (f):} Maps to the Fusion of Paced and Normal beats (denoted as \texttt{f} in the dataset). 
    
    \item \textbf{Fusion of Ventricular and Normal (F):} Maps to the Fusion of Ventricular and Normal beats (denoted as \texttt{F} in the dataset).
\end{itemize}

These five categories (N, A, V, f, F) represent the most critical rhythm types for this study. The mapping preserves the case-sensitivity of the original annotations (e.g., \texttt{f} for paced fusion and \texttt{F} for ventricular fusion) to maintain consistency with the PhysioNet standard.

\subsubsection{Data Preprocessing Pipeline}

Raw ECG recordings typically contain noise and baseline wander. To ensure model robustness and convergence, we implemented a systematic preprocessing pipeline comprising heartbeat segmentation, numerical encoding, signal normalization, and stratified dataset partitioning.

\paragraph{Heartbeat Segmentation}
The continuous ECG recordings were segmented into individual heartbeats to serve as input vectors for the deep learning models. Using the R-peak annotations provided in the dataset as a reference, we extracted a fixed-length temporal window centered on each R-peak. Each extracted heartbeat consists of 188 sampling points (approximately 0.52 seconds), capturing the complete P-QRS-T morphology required for accurate classification.

\paragraph{Label Encoding}
For computational efficiency, the five symbolic class labels (N, A, V, f, F) were converted into numerical integer codes (0-4). This integer representation allows for the efficient computation of categorical cross-entropy loss during the training phase.

\paragraph{Signal Normalization}
To mitigate the impact of amplitude variations across different subjects and recording devices, Min-Max normalization was applied to each heartbeat segment. This process scales the amplitude values to the range $[0, 1]$, ensuring that the neural network weights converge faster and preventing gradients from vanishing or exploding. The normalized signal $x'$ is calculated as:
\begin{equation}
    x' = \frac{x - \min(x)}{\max(x) - \min(x)}
\end{equation}
where $x$ represents the original signal vector.

\paragraph{Dataset Partitioning}
To evaluate the generalization ability of the proposed model, the processed dataset was divided into training and validation subsets. We employed strict stratified sampling to preserve the class distribution in both subsets. Specifically, 85\% of the data was allocated for training and parameter updates, while the remaining 15\% was reserved for validation and model selection.

\subsubsection{Data Augmentation to Balance Data}

Before building the ECG classification model, conducting systematic Exploratory Data Analysis (EDA) on the raw data is a crucial prerequisite for understanding data structure, identifying potential issues, and verifying the effectiveness of preprocessing and augmentation. This study conducts EDA on the MIT-BIH Arrhythmia dataset in four aspects: class distribution, pre- and post-augmentation comparison, waveform visualization, and signal dimension statistics, aiming to provide data evidence for model construction and performance optimization.

To address the severe class imbalance present in arrhythmia datasets, a generative adversarial network (GAN) is employed to synthesize additional samples for underrepresented categories. The GAN framework consists of a generator ($G$) and a discriminator ($D$), where $G$ learns to produce realistic ECG beats while $D$ distinguishes real from synthetic signals. Their objectives are optimized through adversarial learning formulated as follows:

\begin{equation}
\min_{G} \max_{D} \; 
\mathbb{E}_{x \sim p_{\text{real}}}[\log D(x)] 
+ 
\mathbb{E}_{\tilde{x} \sim p_{G}}[\log (1 - D(\tilde{x}))].
\end{equation}

where $\tilde{x} = G(z)$ and $z$ denotes the generator input sequence.

Both the generator ($G$) and the discriminator ($D$) incorporate bidirectional LSTM layers to capture temporal dependencies inherent in ECG waveforms. The generator transforms input noise sequences into realistic one-dimensional ECG beats of 187 samples through fully connected layers with LeakyReLU activations and dropout regularization. The discriminator processes input sequences via a similar LSTM-based architecture, followed by dense layers and a sigmoid output to estimate authenticity. This sequence-aware design ensures that generated beats preserve the temporal continuity and morphological characteristics of real ECG signals, thereby preventing the production of physiologically implausible waveforms.

During training, the GAN is exclusively trained on minority-class samples to learn their morphological variability. Once convergence is achieved, the generator produces synthetic beats that are filtered based on discriminator confidence to ensure sample quality. These high-confidence synthetic beats are then merged with the real training data until the desired class balance is reached. The augmented dataset is subsequently used for model training, allowing the classifiers to learn from a more uniform distribution without altering the overall signal characteristics.

By applying GAN-based data augmentation, the dataset achieves improved balance across arrhythmia classes while preserving physiological fidelity. This approach effectively reduces overfitting to dominant categories and enhances the representativeness of rare cardiac patterns, enabling more reliable and equitable model training across all rhythm types.

Before augmentation, the original dataset exhibits a pronounced class imbalance, with the Normal category significantly outnumbering all others. After applying the generative adversarial network (GAN) for data augmentation, the minority classes are synthetically expanded, resulting in a substantially more balanced distribution. The post-augmentation chart clearly shows a more uniform distribution of samples across categories. For clarity, the bar chart displays both the number of samples and their corresponding percentage of the total for each class.

\subsubsection{ECG Waveform Visualization}

Different types of cardiac rhythms exhibit significant differences in waveform characteristics. To illustrate these differences, five representative samples from each class were randomly selected from the dataset and visualized for comparative analysis. The x-axis corresponds to sampling time points, while the y-axis represents ECG amplitude.

\begin{figure}[htbp]
    \centering
    \includegraphics[width=\textwidth]{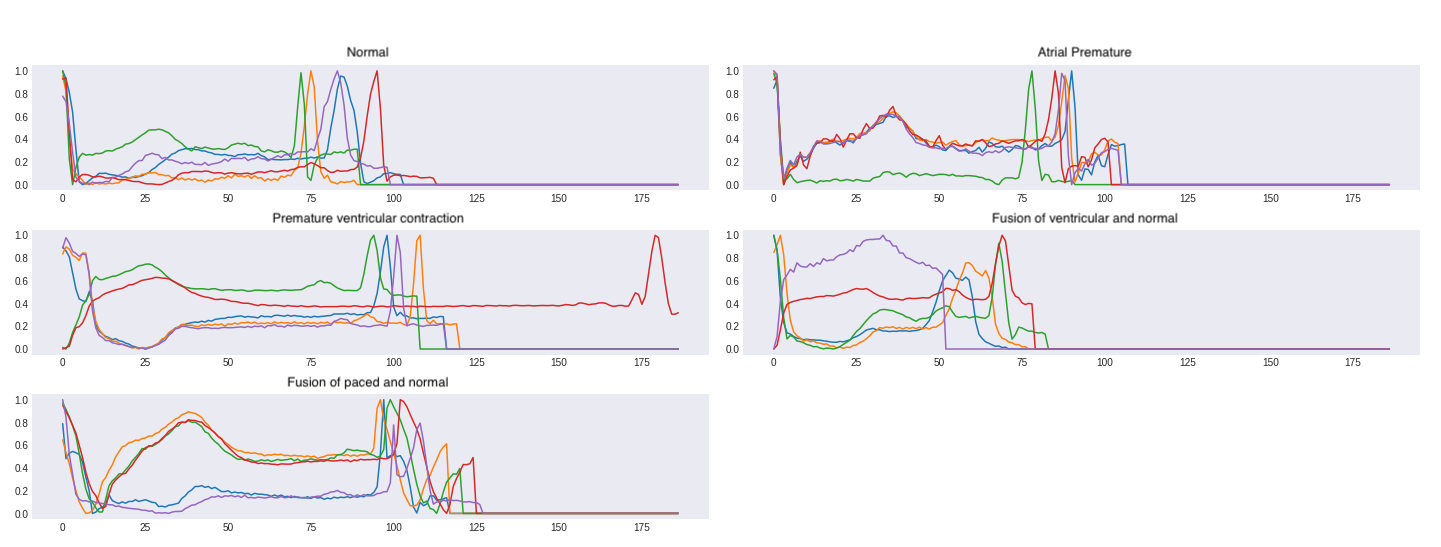}
    \caption{Time-domain waveform examples of the five heartbeat classes: Normal (N), Atrial Premature (A), Premature Ventricular (V), Fusion of Paced and Normal (f), and Fusion of Ventricular and Normal (F).}
    \label{fig:ecg_waveform}
\end{figure}

As shown in Figure~\ref{fig:ecg_waveform}, distinct morphological features can be observed among different arrhythmia types, particularly in critical waveform components such as the QRS complex and T wave. For example, the Premature Ventricular Contraction (V) category typically shows abnormally wide QRS complexes, while the Fusion categories (f and F) exhibit characteristics of mixed rhythms.

\subsection{Model Architectures}

The four deep learning models developed in this study are designed specifically for comparative experiments to evaluate different architectural approaches for ECG classification. Each model represents a distinct strategy for capturing spatial and temporal features from ECG signals. The CNN \cite{kiranyaz2016patient} focuses on extracting local morphological patterns through convolutional operations, while the CNN-LSTM \cite{li2020hybrid} extends this capability by integrating sequential modeling to capture temporal dependencies between consecutive heartbeats. The CNN-LSTM-Attention model further enhances this framework by introducing an attention mechanism that adaptively emphasizes diagnostically important waveform segments, improving interpretability and sensitivity to subtle abnormalities. In contrast, the ResNet-1D \cite{he2016resnet} architecture leverages residual connections to facilitate deeper feature learning and stable gradient propagation, enabling hierarchical representation of complex signal structures.

To leverage the complementary strengths of these architectures, an ensemble learning strategy is subsequently applied, integrating the outputs of all four models to enhance robustness, class balance, and generalization. Collectively, this framework provides a comprehensive foundation for assessing how architectural design and model fusion jointly influence ECG signal representation and classification performance.

\subsubsection{CNN Model}

The proposed CNN model \cite{acharya2017cbm} is a hierarchical one-dimensional convolutional network tailored for sequential signal classification, such as ECG analysis. It comprises three stacked ConvNormPool modules, each performing convolution, normalization, Swish activation, residual connections, and max-pooling. The first module operates directly on the raw ECG signal, while subsequent modules progressively reduce hidden dimensions, capturing multi-scale temporal features.

Each convolutional operation extracts local morphological patterns from the ECG waveform, formally defined as:
\begin{equation}
    h_i = \sigma(W * x_i + b)
\end{equation}
where $W$ and $b$ denote the convolution kernel and bias, respectively, and $\sigma(\cdot)$ represents the Swish activation function:
\begin{equation}
    \sigma(x) = x \cdot \text{sigmoid}(x)
\end{equation}

This non-linear function enhances expressivity while maintaining smooth gradient propagation.

Each ConvNormPool \cite{gu2015cnnreview} block integrates normalization layers to stabilize training, as well as skip connections to preserve low-level representations and facilitate efficient gradient flow. Max-pooling layers reduce temporal resolution, allowing the network to focus on salient waveform patterns at multiple scales.

Following the convolutional stages, an adaptive average pooling layer compresses the extracted features into a fixed-length representation, which is passed to a final fully connected layer to compute class probabilities. This architecture balances computational efficiency with representational power, enabling real-time and accurate classification of ECG signals by effectively capturing both local morphology and intermediate temporal dependencies.

\begin{figure}[htbp]
    \centering
    \includegraphics[width=0.9\textwidth]{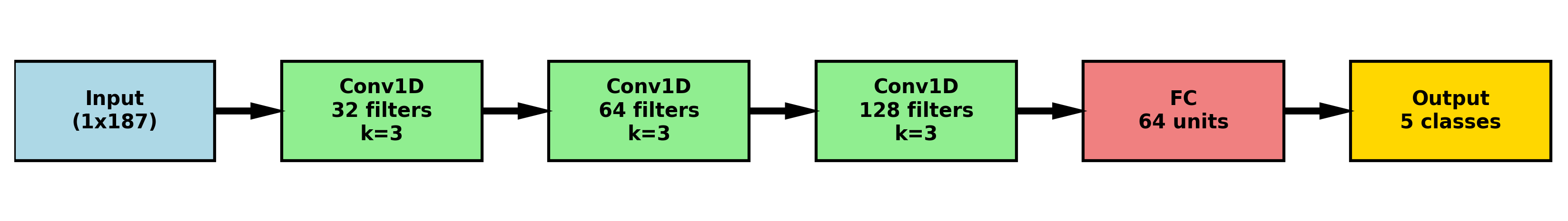} 
    \caption{CNN Model Architecture}
    \label{fig:cnn_model}
\end{figure}

\subsubsection{CNN-LSTM Model}

The proposed CNN-LSTM model integrates convolutional and recurrent layers to capture both local and long-term dependencies in sequential signals. The architecture begins with two ConvNormPool modules, which apply convolution, normalization, Swish activation, and residual connections to extract multi-scale temporal features. Max-pooling operations progressively reduce sequence length while emphasizing salient patterns.

The extracted representations are subsequently processed by a bidirectional Long Short-Term Memory (LSTM) \cite{hochreiter1997lstm} layer that models sequential dependencies across time steps. The internal operations of the LSTM cell are governed by gating mechanisms that regulate information flow, formulated as:
\begin{align}
    h_t &= o_t \odot \tanh(C_t) \\
    C_t &= f_t \odot C_{t-1} + i_t \odot \widetilde{C}_t
\end{align}
where $f_t$, $i_t$, and $o_t$ denote the forget, input, and output gates, respectively, and $\odot$ indicates element-wise multiplication. This recurrent structure enables the model to preserve long-term temporal context, capturing rhythmic variations between consecutive ECG beats.

This recurrent layer supports multi-layer configurations and bidirectionality, enabling the network to capture both forward and backward contextual information. Dropout is included to improve generalization and prevent overfitting.

After the recurrent stage, adaptive average pooling compresses the temporal dimension into a fixed-size vector, which is then processed by a fully connected layer to produce class probabilities. By combining convolutional feature extraction with recurrent temporal modeling, the CNN-LSTM architecture effectively balances local pattern recognition and sequence-level dependency learning, making it well-suited for tasks such as ECG classification and time-series analysis.

\begin{figure}[htbp]
    \centering
    \includegraphics[width=0.9\textwidth]{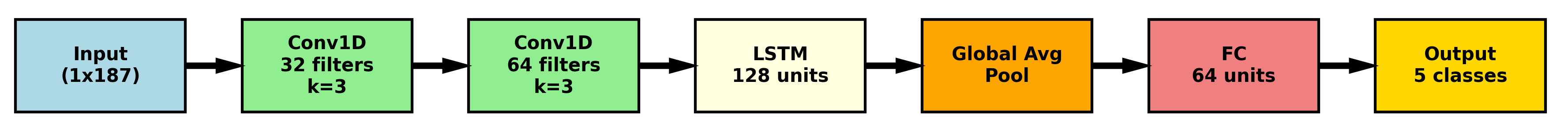}
    \caption{CNN-LSTM Model Architecture}
    \label{fig:cnn_lstm_model}
\end{figure}

\subsubsection{CNN-LSTM-Attention Model}

The CNN-LSTM-Attention model extends the CNN-LSTM hybrid architecture by incorporating an attention mechanism that adaptively emphasizes diagnostically relevant temporal regions of the ECG signal. The model begins with two ConvNormPool modules, which extract hierarchical temporal features through convolution, normalization, Swish activation, and residual connections. These layers capture local waveform dependencies while reducing sequence length and highlighting salient patterns.

The intermediate feature representations are then processed by a bidirectional Long Short-Term Memory (LSTM) layer to model sequential dependencies in both temporal directions. Bidirectional processing enables the network to capture contextual information from both past and future time steps, improving rhythm recognition and long-range temporal coherence\cite{hannun2019deep} .

To further enhance interpretability and focus, an attention layer is applied on top of the LSTM outputs to assign varying importance to different time segments. The attention \cite{vaswani2017attention}  weights are computed as:
\begin{equation}
    \alpha_t = \text{softmax}\left( v^{\top} \tanh(W_h h_t + b_h) \right)
\end{equation}
where $h_t$ denotes the hidden state at time step $t$, $W_h$ and $v$ are trainable parameters, and $\alpha_t$ represents the relative importance of each temporal feature. The final context vector is obtained as:
\begin{equation}
    s = \sum_t \alpha_t h_t
\end{equation}
which aggregates the most informative temporal features into a compact representation.

Subsequently, adaptive max-pooling compresses the attended features into a fixed-length vector, which is mapped to class probabilities by a fully connected layer. By unifying convolutional feature extraction, recurrent temporal modeling, and attention-based feature weighting, the CNN-LSTM-Attention model achieves robust performance on complex sequential classification tasks such as ECG analysis.

\begin{figure}[ht]
    \centering
    \includegraphics[width=0.9\textwidth]{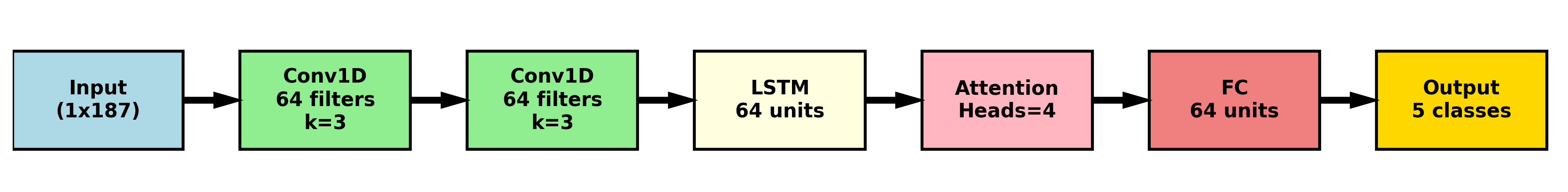}
    \caption{CNN-LSTM-Attention Model Architecture}
    \label{fig:cnn_lstm_attention}
\end{figure}

\subsubsection{ResNet-1D Model}

The ResNet-1D model adapts the residual learning framework of the original ResNet architecture, tailored for one-dimensional temporal signal classification such as ECG analysis. It begins with a 1D convolutional layer (kernel size 7, stride 2), followed by batch normalization, ReLU activation, and max-pooling to extract low-level features and reduce sequence length \cite{gu2015cnnreview}.

The core of the architecture consists of multiple residual blocks, each containing two 1D convolutional layers with batch normalization and ReLU activations. A shortcut (identity) connection bypasses each block to directly propagate the input to the output, mitigating gradient vanishing and improving optimization stability. The forward propagation of a residual block can be expressed as:
\begin{equation}
    y = F(x, \{W_i\}) + x
\end{equation}
where $x$ is the input, $F(x, \{W_i\})$ denotes the residual mapping implemented by successive convolutional layers with parameters $\{W_i\}$, and $y$ represents the block output. When the input and output dimensions differ, a $1 \times 1$ convolution is applied to match channel dimensions before the addition.

This residual formulation allows the network to learn hierarchical representations efficiently, enabling deeper feature extraction without degradation in training accuracy. After the final residual stage, adaptive average pooling aggregates temporal features into a fixed-length vector, which is passed through a fully connected layer and softmax activation to produce class probabilities.

By combining hierarchical feature learning with stable gradient propagation, the ResNet-1D architecture effectively captures both local and global temporal dependencies in ECG signals. Its strong generalization and robustness make it particularly suitable for deep arrhythmia classification tasks.

\begin{figure}[ht]
    \centering
    \includegraphics[width=0.9\textwidth]{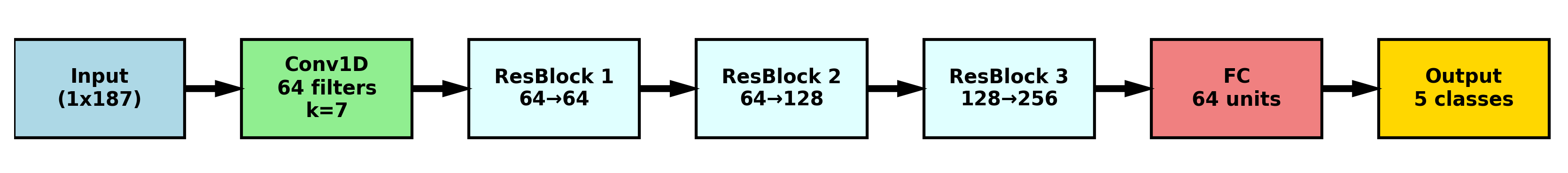}
    \caption{ResNet-1D Model Architecture}
    \label{fig:resnet1d}
\end{figure}

\subsection{Ensemble Learning Strategy}

To improve robustness and class balance beyond any single architecture, multiple trained networks were aggregated at inference time through a linear late-fusion of their outputs. Let $f_m(x) \in \mathbb{R}^C$ denote the class-logit vector produced by model $m$ for an ECG segment $x$, where $m = 1,\dots,M$ and $C$ represents the number of classes. The ensemble prediction is obtained by a weighted average of model outputs followed by an argmax operation:
\begin{equation}
    \hat{z}(x) = \sum_{m=1}^{M} w_m f_m(x)
\end{equation}
\begin{equation}
    \sum_{m=1}^{M} w_m = 1, \quad w_m \ge 0
\end{equation}
\begin{equation}
    \hat{y}(x) = \arg\max_c \hat{z}_c(x)
\end{equation}

This late-fusion strategy is architecture-agnostic, avoids retraining, and exploits complementary error patterns observed in confusion matrix and ROC analyses, particularly the trade-offs between dominant rhythms and minority arrhythmias.

The \texttt{EnsembleModel} module was implemented as a wrapper around a list of trained base learners, computing a convex combination of their output logits. When no explicit weights were specified, uniform averaging was applied and automatically normalized to ensure $\sum w_m = 1$. Evaluation followed the same protocol as individual models, reporting accuracy, macro-F1, macro-precision, and macro-recall on the held-out test set.

Four practical fusion strategies were explored to probe the bias-variance trade-off and complementary model contributions:
\begin{enumerate}
    \item \textbf{All-Equal-Weight:} all models aggregated equally.
    \item \textbf{Top3-Equal-Weight:} the three best-performing models averaged equally.
    \item \textbf{Top2-Equal-Weight:} the two best models averaged equally.
    \item \textbf{Top2-Weighted:} weighted fusion of the top two models based on their validation macro-F1 scores.
\end{enumerate}

If the best two models have validation macro-F1 scores $F_1^{(1)}$ and $F_1^{(2)}$, the fusion weights are defined as:
\begin{equation}
    w_1 = \frac{F_1^{(1)}}{F_1^{(1)} + F_1^{(2)}}, \quad 
    w_2 = \frac{F_1^{(2)}}{F_1^{(1)} + F_1^{(2)}}
\end{equation}

This formulation prioritizes the stronger model while maintaining contributions from the complementary one. It is particularly beneficial when one model excels on dominant classes while another contributes improved sensitivity for rare arrhythmias. Averaging logits reduces variance and mitigates idiosyncratic errors from individual networks. The top-$k$ and weighted variants balance diversity and strength, emphasizing models with consistent F1 superiority.

For each fusion strategy, predictions on the test set were generated using the same data loader as individual models. Performance metrics including accuracy, macro-F1, macro-precision, and macro-recall were computed. When multiple runs were available, confidence intervals were calculated following the same methodology used in previous CI analyses.

The proposed late-fusion ensemble framework is straightforward, computationally efficient, and empirically effective. It enhances classification stability and class balance across heterogeneous ECG morphologies, providing a principled extension beyond individual model performance.

\section{Experimental Results}

This section presents the experimental setup, training configuration, and evaluation procedures used to validate the proposed architectures. It further compares the quantitative and qualitative performance of all models and analyzes the impact of ensemble learning on classification robustness and generalization.
\subsection{Experimental Environment}

All experiments were conducted in PyTorch, with training and inference performed on an NVIDIA Tesla T4 GPU (16 GB memory) to ensure consistent computational efficiency. 

The datasets used in this study are the well-known MIT-BIH Arrhythmia Database and the PTB Diagnostic ECG Database, which were fully introduced in Section~\ref{sec:dataset_intro} of this paper.

A standardized training and optimization framework was adopted across all architectures, incorporating systematic hyperparameter tuning through grid search and empirical adjustment. Each model was trained using identical preprocessing pipelines, data splits, and optimization settings to ensure a fair comparison across architectures. 
The key hyperparameter configurations (e.g., batch size, learning rate, optimizer, and loss function) were standardized and reported in Section~\ref{sec:hyperparams}. 

All experiments were conducted under controlled and consistent settings to ensure reproducibility, using the same data split and preprocessing pipeline for all architectures to allow fair performance comparison.

This controlled experimental setup ensures that performance differences across CNN, CNN-LSTM, CNN-LSTM-Attention, ResNet-1D, and their ensemble variants reflect architectural effectiveness rather than external factors such as data imbalance or random initialization.

\subsection{Training Configuration}

To ensure stable convergence and class-balanced learning, all models were trained under a standardized optimization framework consisting of Focal Loss, the AdamW optimizer, and the ReduceLROnPlateau scheduler. These components jointly enhance minority-class sensitivity and prevent overfitting under highly imbalanced ECG data distributions.

\subsubsection{Focal Loss}

Given the imbalanced distribution of the ECG dataset, where Normal heartbeats dominate while critical arrhythmia classes such as Fusion of ventricular and normal and Atrial Premature are underrepresented, conventional loss functions often bias toward the majority class. To mitigate this, Focal Loss \cite{lin2017focalloss} is employed, which adaptively scales the loss contribution of each sample according to the difficulty of classification, effectively emphasizing minority and hard-to-classify examples. The standard cross-entropy loss tends to be dominated by easy and majority-class samples, leading to suboptimal learning for minority classes. Focal Loss introduces a modulating factor that dynamically down-weights well-classified examples and focuses training on hard, misclassified samples.

Formally, the Focal Loss is defined as:

\begin{equation}
\mathrm{FL}(p_t) = -\alpha_t (1 - p_t)^{\gamma} \log(p_t)
\end{equation}

where \( p_t \) denotes the predicted probability for the true class, 
\( \alpha_t \) is a weighting factor that balances the importance of different classes, 
and \( \gamma \) is a focusing parameter that adjusts the rate at which easy examples are down-weighted.

When \( \gamma = 0 \), Focal Loss reduces to standard cross-entropy. As \( \gamma \) increases, the relative loss for well-classified examples (those with large \( p_t \)) decreases, allowing the model to pay more attention to misclassified or minority-class samples. This mechanism effectively mitigates the dominance of majority classes during training, improving model robustness under class imbalance.

In this work, the Focal Loss is implemented with \( \alpha = 1 \) and \( \gamma = 2 \), which provides a balance between convergence stability and minority-class sensitivity. Compared with conventional weighted cross-entropy, Focal Loss offers smoother gradient adjustment, which was empirically more stable for rare arrhythmia classes in this study.

\subsubsection{AdamW Optimizer}
We employed the AdamW optimizer \cite{loshchilov2019adamw}, which decouples weight decay from the gradient-based parameter updates, thereby improving generalization compared to the conventional Adam optimizer. Unlike standard Adam, where \( L_2 \) regularization is implicitly intertwined with the adaptive moment estimation, AdamW explicitly separates the weight decay term from the gradient updates. This modification leads to a more principled regularization effect and prevents the optimizer from overfitting to the training data.

Moreover, AdamW preserves the adaptive learning rate and momentum advantages of Adam while offering better convergence stability and faster training across a wide range of tasks. This makes it particularly effective for deep neural networks with large parameter spaces, such as transformer-based and convolutional architectures. Compared with traditional optimizers such as SGD with momentum or RMSProp, AdamW requires less manual tuning of the learning rate and exhibits robust performance under class imbalance and noisy data conditions, which are common challenges in ECG signal classification. Therefore, AdamW provides a balanced trade-off between convergence efficiency and generalization capability, resulting in more stable and reliable training outcomes in our experiments.

\subsubsection{ReduceLROnPlateau Learning Rate Scheduler}
To further stabilize the training process and enhance convergence, we employed the ReduceLROnPlateau learning rate scheduler. This scheduler dynamically adjusts the learning rate based on the validation loss, reducing it when the metric stagnates over a predefined number of epochs. Specifically, if the validation loss fails to decrease after a set patience period, the scheduler scales the learning rate by a specified factor. This adaptive adjustment mechanism helps prevent the optimizer from getting trapped in local minima and promotes smoother convergence toward the global optimum.

Formally, the update rule can be expressed as:

\begin{equation}
\eta_{t+1} = \max(\eta_t \cdot \text{factor}, \text{min\_lr})
\end{equation}

where the parameter \(\text{factor}\) (\(0 < \text{factor} < 1\)) determines the magnitude of reduction, and \(\text{min\_lr}\) sets the lower learning-rate bound. In this work, the scheduler monitored the validation loss in \texttt{"min"} mode with \(\text{factor} = 0.5\) and \(\text{patience} = 3\), meaning that the learning rate was halved when the loss failed to improve for three consecutive epochs. This event-driven policy allows the model to maintain a higher learning rate when useful progress continues and automatically transition to a finer optimization phase once improvement slows.

Compared with fixed or cosine-annealed schedules, the \texttt{ReduceLROnPlateau} strategy provides several advantages for ECG classification tasks. It adapts to the irregular learning patterns that arise from imbalanced data and inter-patient variability, effectively filtering transient fluctuations through its patience threshold. The scheduler promotes smoother convergence, prevents overfitting during late-stage training, and requires minimal manual tuning, offering consistency across different architectures and datasets.

In our implementation, the scheduler operated jointly with the AdamW optimizer (\(\text{weight\_decay} = 1 \times 10^{-4}\)), and the validation loss served as the monitored metric. This configuration ensured that learning rate reductions occurred before early stopping was triggered, allowing the network to recover from temporary plateaus and achieve improved generalization performance across diverse ECG morphologies.

\subsubsection{Hyperparameter Setting}
\label{sec:hyperparams}

This study uses grid search and empirical tuning across multiple experiments to determine the optimal combination of key hyperparameters:

\begin{table}[htbp]
\centering
\resizebox{\textwidth}{!}{
\begin{tabular}{lccccccc}
\hline
Model & Batch Size & Initial Learning Rate & Patience & Optimizer & Loss & Epoch & Scheduler \\
\hline
CNN & 128 & $1.15\times10^{-3}$ & 8 & AdamW & FocalLoss & 50 & ReduceLROnPlateau \\
CNN-LSTM & 96 & $1\times10^{-3}$ & 8 & AdamW & FocalLoss & 50 & ReduceLROnPlateau \\
CNN-LSTM-Attention & 96 & $1\times10^{-3}$ & 8 & AdamW & FocalLoss & 50 & ReduceLROnPlateau \\
ResNet-1D & 96 & $1.22\times10^{-3}$ & 8 & AdamW & FocalLoss & 50 & ReduceLROnPlateau \\
\hline
\end{tabular}}
\caption{Training configurations for all models.}
\end{table}

\subsection{Evaluation Metrics and Interpretability}

Model performance was evaluated through a combination of quantitative metrics and qualitative interpretability analyses to comprehensively assess classification accuracy, class balance, and model reliability.

\subsubsection{Quantitative Evaluation Metrics}

To quantify classification performance, the study adopted standard evaluation metrics, including Accuracy, Precision, Recall, and F1-score, along with Receiver Operating Characteristic (ROC) and Area Under the Curve (AUC) analyses. Accuracy measures the overall proportion of correctly classified samples, while Precision and Recall capture the trade-off between false positives and false negatives. The F1-score, defined as the harmonic mean of Precision and Recall, provides a balanced measure of sensitivity and specificity, making it particularly suitable for the imbalanced class distribution characteristic of ECG datasets.

A Confusion Matrix was also employed to visualize class-wise prediction outcomes. Diagonal elements represent correctly predicted samples, whereas off-diagonal elements highlight common misclassifications between morphologically similar rhythms (e.g., Normal vs.\ Atrial Premature Beats). This visualization offers detailed insights into class-specific performance and error patterns.


This threshold-independent measure is particularly informative for imbalanced datasets, offering a robust comparison across models and supporting the identification of clinically relevant decision thresholds. The ROC curve illustrates the trade-off between sensitivity and specificity across different decision thresholds, while the AUC provides a scalar summary of the model’s overall separability.

For ROC analysis, curves were first generated for each arrhythmia class across all models to compare class-wise discrimination performance. Subsequently, macro-average AUC values were computed by averaging the per-class results, providing an overall model assessment that ensures equal weighting across all arrhythmia categories.

\subsubsection{Statistical Confidence Analysis}

To assess the reliability and statistical robustness of the reported results, 95\% confidence intervals (CIs) were computed for the key evaluation metrics. CIs provide a statistical measure of uncertainty by defining a range within which the true population metric is expected to lie. Narrower intervals indicate greater stability and reproducibility, whereas wider intervals reflect higher variability in model performance. Reporting confidence intervals alongside point estimates ensures transparency and offers a statistically grounded interpretation of the results.

\subsubsection{Interpretability and Visualization}

Beyond numerical evaluation, model interpretability was further examined using Gradient-weighted Class Activation Mapping (Grad-CAM), a visualization technique that leverages the gradients flowing into the final convolutional layer to produce class-discriminative activation maps. Grad-CAM highlights the temporal regions of ECG signals that most strongly influence model predictions, revealing which waveform segments (e.g., QRS complexes, P-waves, or premature contractions) contribute most to classification decisions.

These visual explanations help verify whether the models attend to physiologically meaningful cardiac features, thereby enhancing interpretability and clinical trustworthiness. Together, these evaluation metrics and interpretability analyses provide a comprehensive understanding of both quantitative performance and qualitative reasoning, ensuring that each model’s assessment reflects not only predictive accuracy but also interpretability and clinical relevance. The following section presents the quantitative results and comparative analysis across all evaluated architectures.

\subsection{Model Performance Comparison}

\subsubsection{Training and Validation Performance Comparison}
\label{sec:train_val_comparison}

\begin{figure}[!htbp]
\centering

{\scriptsize (a) CNN}\\[-0.3em]
\includegraphics[width=0.8\textwidth]{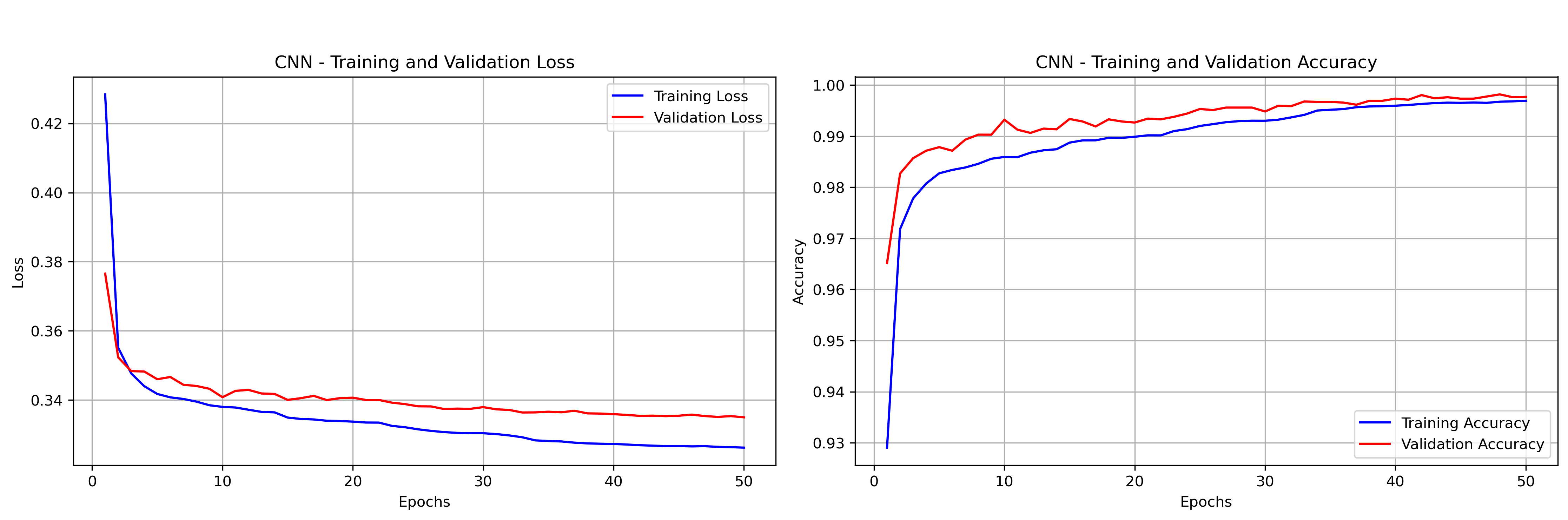}

\vspace{1em}

{\scriptsize (b) CNN-LSTM}\\[-0.3em]
\includegraphics[width=0.8\textwidth]{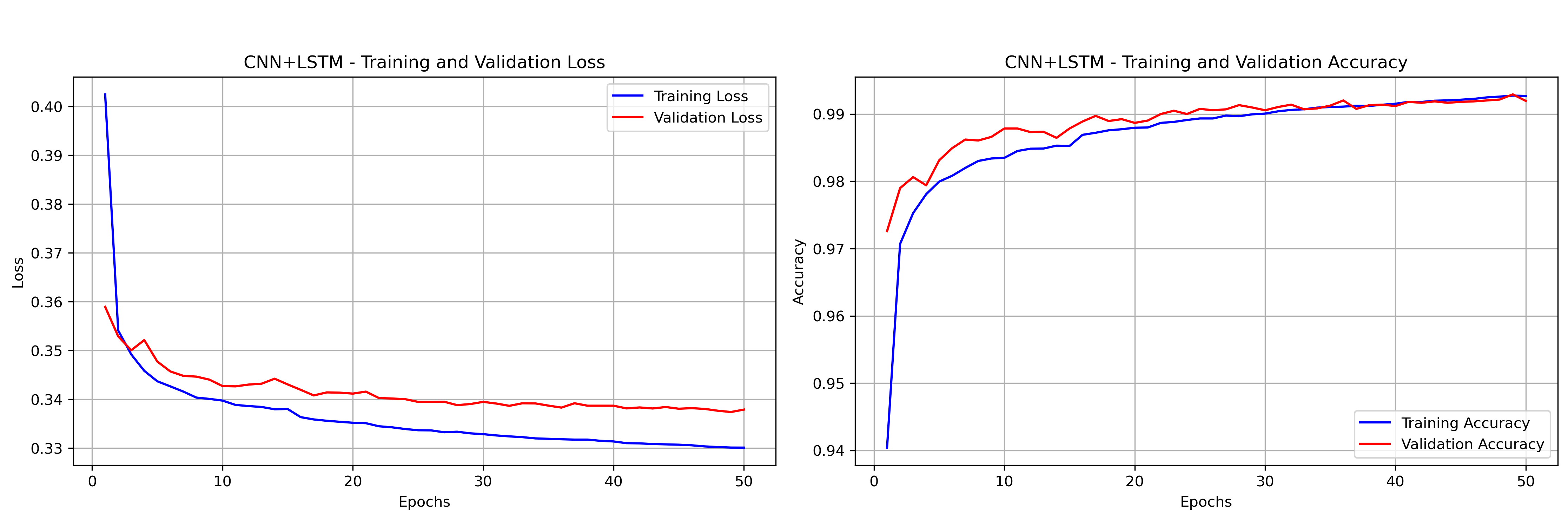}

\vspace{1em}

{\scriptsize (c) CNN-LSTM-Attention}\\[-0.3em]
\includegraphics[width=0.8\textwidth]{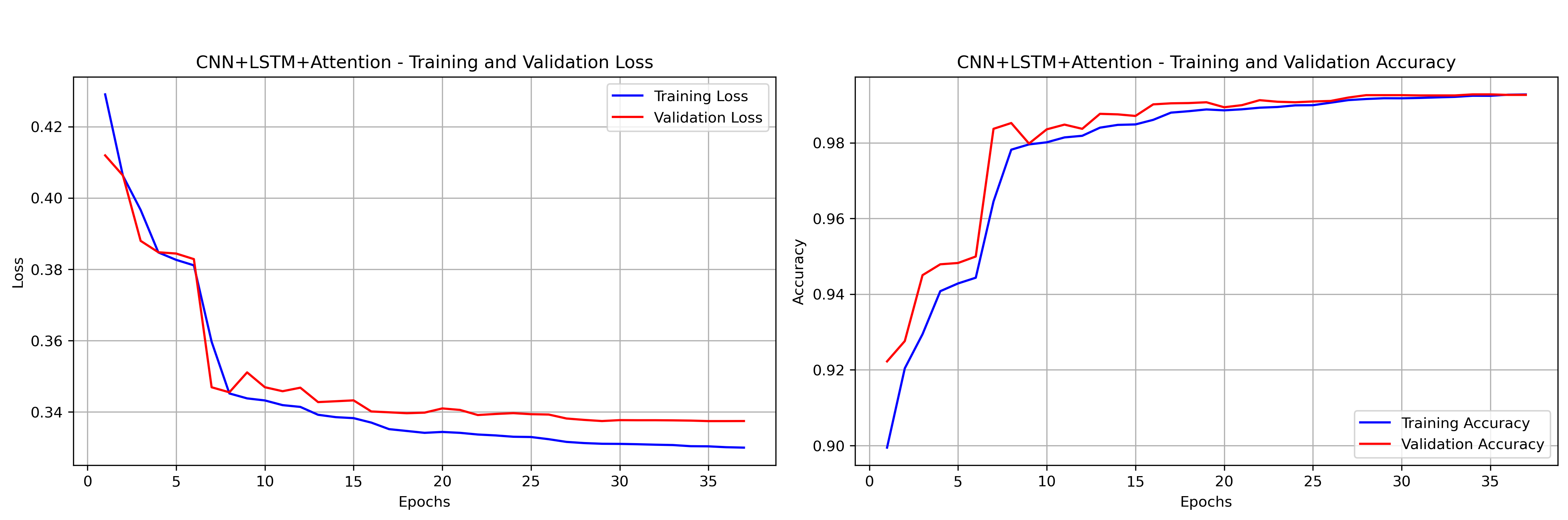}

\vspace{1em}

{\scriptsize (d) ResNet}\\[-0.3em]
\includegraphics[width=0.8\textwidth]{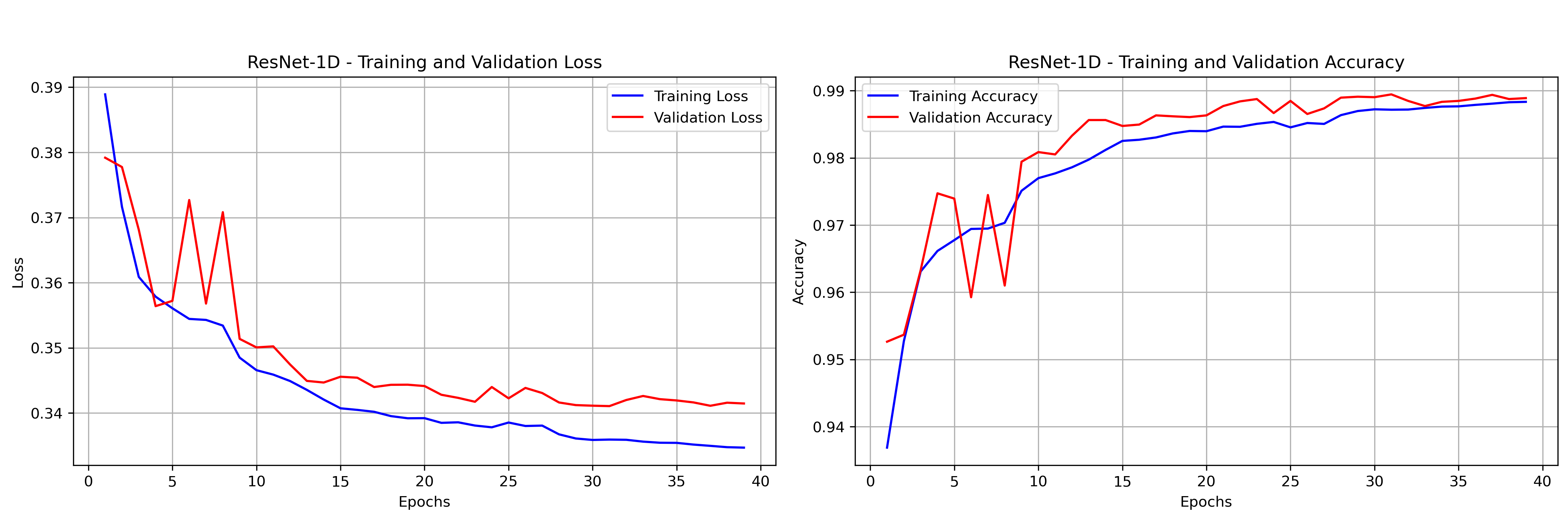}

\caption{Training and validation curves for all models.}
\label{fig:training_curves}
\end{figure}

The training and validation loss curves of the four models are presented in Figure~\ref{fig:training_curves} to analyze the convergence characteristics, optimization stability, and generalization behavior during ECG classification. The CNN model demonstrates a clear and smooth downward trend in both training and validation losses. The loss decreases sharply within the first few epochs, then gradually stabilizes as the network converges. Although the validation loss consistently remains higher than the training loss, the two curves follow similar trajectories, indicating mild overfitting but stable learning dynamics. The validation accuracy also closely tracks the training accuracy throughout training, further confirming that the model learns consistently without significant divergence. These results suggest that the CNN effectively captures local morphological patterns such as QRS complexes and P-wave structures, enabling reliable convergence in ECG classification.

Compared with the purely convolutional model, the CNN-LSTM architecture demonstrates more synchronized training dynamics. Both training and validation losses decrease rapidly in the early epochs, and the two curves remain tightly aligned throughout the entire training process, with a consistently smaller gap than that observed in the CNN model. The validation loss also exhibits reduced fluctuation, indicating smoother and more stable optimization.

A similar pattern appears in the accuracy curves: training and validation accuracy rise in near-parallel trajectories and converge closely across epochs, with the validation accuracy maintaining a slight but stable lead. The CNN-LSTM model shows more tightly coupled and smoother learning behavior than the baseline CNN, suggesting that incorporating temporal recurrence contributes to steadier gradient updates and improved generalization.

The CNN-LSTM-Attention architecture converges rapidly during the initial epochs but shows minor oscillations in validation loss. These fluctuations indicate that the additional attention mechanism increases model complexity, which slightly reduces quantitative stability under the current dataset size. While the attention layer improves interpretability by emphasizing diagnostically relevant waveform regions, it offers limited quantitative advantage compared with the CNN-LSTM baseline.

In contrast, the ResNet-1D model converges more slowly and presents noticeable oscillations in the validation loss during early epochs. While its training loss decreases steadily, the validation loss exhibits larger fluctuations before gradually stabilizing. This pattern suggests that the model’s deeper residual structure, although capable of capturing hierarchical features, is more sensitive to parameter initialization and regularization. The wider gap between the training and validation losses indicates a higher risk of overfitting when trained on limited ECG samples. Despite achieving competitive accuracy, its convergence trajectory demonstrates moderate instability compared with the recurrent models.

To complement the visual observations, Table~\ref{tab:qualitative_summary} qualitatively summarizes the convergence behavior, validation stability, and overall performance trends of each model.

\begin{table}[t]
\centering
\small
\setlength{\tabcolsep}{3pt}
\resizebox{\textwidth}{!}{%
\begin{tabular}{p{0.13\textwidth} p{0.16\textwidth} p{0.18\textwidth}
                p{0.20\textwidth} p{0.28\textwidth}}
\hline
\begin{tabular}[c]{@{}l@{}}%
Model
\end{tabular} & 
\begin{tabular}[c]{@{}c@{}}%
Convergence \\ Behavior
\end{tabular} &
\begin{tabular}[c]{@{}c@{}}%
Validation \\ Stability
\end{tabular} &
\begin{tabular}[c]{@{}c@{}}%
Overall \\ Performance 
\end{tabular} &
\begin{tabular}[c]{@{}c@{}}%
Observed \\ Characteristics
\end{tabular} \\
\hline

CNN &
Smooth and steady loss decay &
Stable but mild overfitting &
High accuracy (\(\sim 0.99\)) &
Reliable baseline; strong local feature modeling but no explicit temporal modeling. \\

CNN-LSTM &
Fast and smooth convergence &
Most stable (lowest variance) &
Best balance of convergence, stability, and generalization &
Captures long-term temporal structure; robust and reliable across classes. \\

CNN-LSTM-ATTN &
Rapid early convergence with slight oscillations &
Moderate instability &
Slightly lower quantitative performance &
Attention improves interpretability but increases optimization complexity. \\

ResNet-1D &
Slow convergence with periodic fluctuations &
High variance &
Competitive accuracy &
Deep residual architecture; more sensitive to training dynamics and hyperparameters. \\

\hline
\end{tabular}
}
\caption{Qualitative summary of model convergence, validation stability, and observed training characteristics.}
\label{tab:qualitative_summary}
\setlength{\tabcolsep}{6pt}
\end{table}

Overall, the CNN-LSTM architecture provides the best trade-off between convergence speed, stability, and generalization, whereas the attention-augmented variant adds interpretability at the cost of slightly increased training variability. These findings suggest that different architectures learn complementary aspects of ECG morphology and temporal dynamics. Specifically, CNN captures local waveform features efficiently, LSTM models sequential dependencies, the attention mechanism adaptively highlights diagnostically relevant segments, and ResNet-1D provides deep hierarchical representations. Recognizing these complementary strengths, an ensemble strategy was subsequently developed to integrate their diverse learning behaviors, with the goal of improving robustness, class-wise balance, and overall performance.

\subsubsection{Per-Class Performance Matrix Comparison}

\begin{figure}[!htbp]
\centering

\includegraphics[width=0.6\textwidth]{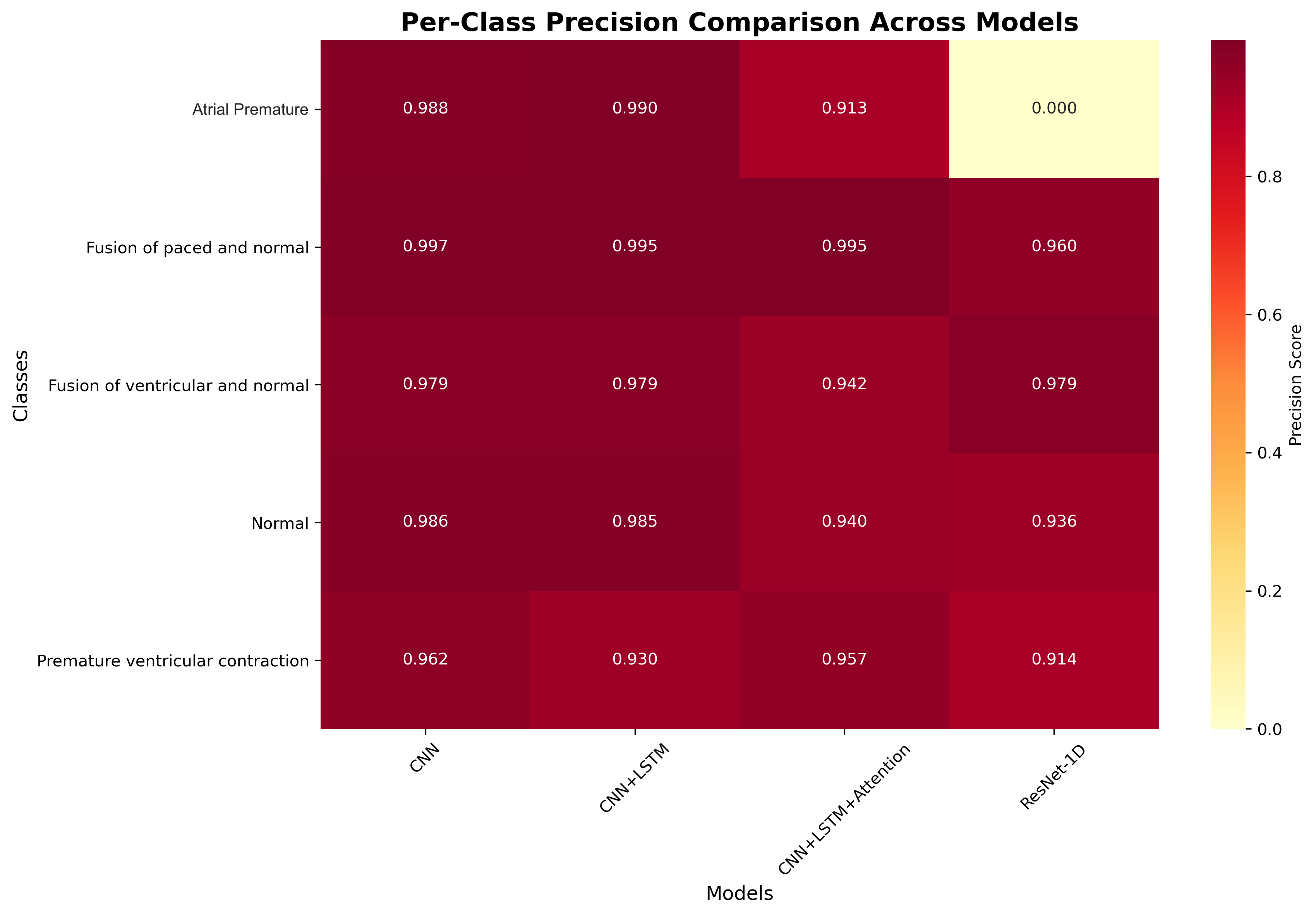}
\hfill
\includegraphics[width=0.6\textwidth]{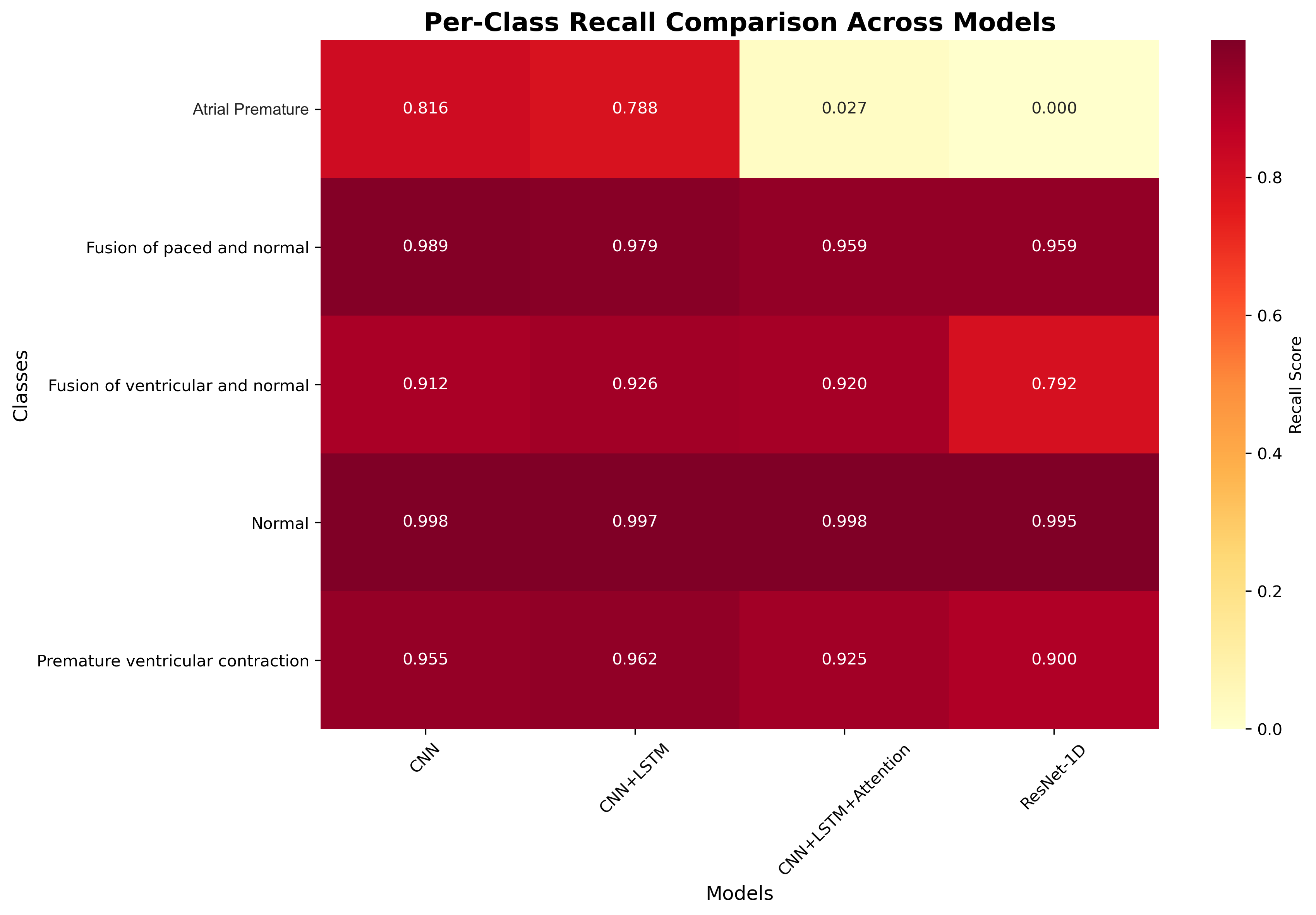}
\vspace{1em}
\includegraphics[width=0.6\textwidth]{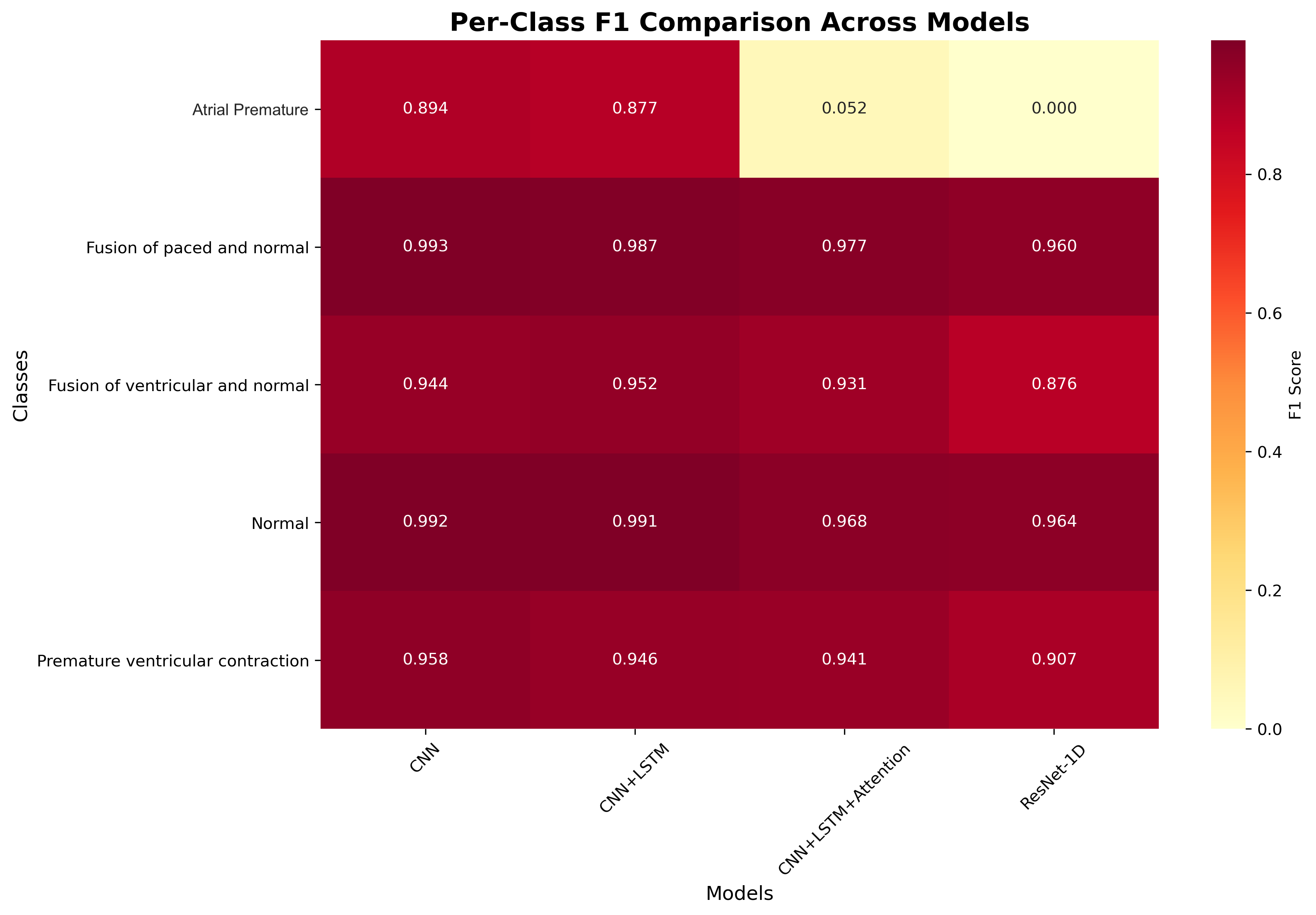}

\caption{Per-class Precision, Recall, and F1-score comparison across models.}
\label{fig:per_class_metrics}
\end{figure}

Figure~\ref{fig:per_class_metrics} presents the per-class Precision, Recall, and F1-score comparison across all evaluated architectures. These heatmaps provide a detailed visualization of each model’s discriminative capability across five ECG rhythm types \cite{aami2008ec57}: Normal (N), Atrial Premature (A), Premature Ventricular Contraction (V), Fusion of Paced and Normal (f), and Fusion of Ventricular and Normal (F), revealing clear differences in sensitivity to class imbalance and inter-class morphological variation.

The trends observed here are consistent with the convergence behaviors discussed in Section~\ref{sec:train_val_comparison}. Models that exhibited stable training and validation loss trajectories, such as CNN-LSTM, also demonstrated consistent generalization across classes. In contrast, architectures with oscillatory or unstable convergence, such as CNN-LSTM-Attention and ResNet-1D, showed more pronounced inter-class performance variance. This coherence between optimization stability and classification outcomes highlights the reliability of convergence patterns as indicators of generalization robustness in ECG analysis.

Across all models, Normal beats were recognized with near-perfect recall (\(\approx 0.99\)), reflecting their distinctive morphology and high prevalence in the dataset. Fusion-type beats (f and F) also achieved consistently strong performance across most architectures, although certain models showed noticeable degradation, particularly ResNet-1D for F and the Attention-based model. In contrast, the classification of minority classes, especially Atrial Premature (A) and Ventricular Premature (V), diverged markedly across models, revealing substantial sensitivity to architectural design and class imbalance.

The CNN baseline achieved strong and stable results overall, with precision above 0.96 across all categories and recall remaining high for most classes, while showing moderate degradation for minority arrhythmias. It achieved the highest recall (0.82) for the Atrial Premature (A) class, indicating that convolutional filters effectively capture localized morphological irregularities such as atypical QRS complexes and P-wave distortions. However, this advantage is class-specific rather than global, as the CNN’s fixed and limited temporal receptive field restricts its ability to model long-range dependencies across consecutive beats. As a result, although the CNN learns local waveform morphology efficiently, its overall generalization across classes that benefit from broader temporal context is weaker than that of recurrent architectures, which are explicitly designed to integrate beat-to-beat temporal information.

The CNN-LSTM model exhibited slightly lower recall for A-type beats (0.79) but maintained the most balanced class-wise performance overall. It attained high recall (\(\ge 0.96\)) for Normal, Fusion of Paced and Normal (f), and Ventricular Premature (V) beats, which together constitute the majority portion of the dataset. Its recurrent component effectively captured temporal continuity and improved overall class-wise balance without introducing overfitting, confirming that temporal modeling enhances generalization stability. These results highlight a trade-off between localized morphological precision and broader temporal modeling: while the CNN is more sensitive to localized waveform irregularities, the CNN-LSTM achieves more balanced performance across rhythm types by integrating temporal dependencies throughout the signal.

In contrast, the CNN-LSTM-Attention model performed reasonably well on the majority classes, with precision remaining high (approximately 0.91–0.99) and recall above 0.92 for most of them. However, it collapsed on the Atrial Premature (A) class (recall \(\approx 0.03\), \(F_1 \approx 0.05\)). This degradation suggests that although the attention mechanism can enhance feature localization for prevalent rhythms, it amplifies class-imbalance effects and loses robustness when rare patterns are underrepresented. This indicates that attention-based architectures may require larger or more balanced datasets to achieve stable generalization across rhythm types.

The ResNet-1D model achieved high but inconsistent precision for the frequent classes, ranging from approximately 0.91 to 0.96, and exhibited the widest recall variance among all architectures, spanning from 0.99 for Normal beats to 0.79 for F and 0.00 for A-type beats. These fluctuations align with the convergence oscillations observed earlier, indicating that the deeper residual structure increases sensitivity to initialization and encourages overfitting toward majority-class patterns under imbalanced data conditions.

From a global perspective, the CNN-LSTM achieved the best macro-average performance (Precision = 0.972, Recall = 0.930, \(F_1 = 0.955\)), effectively balancing sensitivity and specificity. The CNN remained a strong morphological baseline with superior recall for A-type arrhythmias, reflecting its strength in capturing localized waveform irregularities. In contrast, the CNN-LSTM-Attention and ResNet-1D models demonstrated substantial instability under pronounced class imbalance, leading to severe degradation in minority-class performance.

Heatmaps across five arrhythmia classes reveal that the CNN-LSTM model provides the most balanced class-wise performance, whereas the attention-based and residual variants exhibit severe recall degradation on minority categories, particularly Atrial Premature (A). These results indicate that incorporating temporal modeling via recurrent layers enhances overall generalization stability and reduces class-wise variability, especially under strong class imbalance.

\subsubsection{Confusion Matrix Comparison}

\begin{figure}[htbp]
    \centering
    \includegraphics[width=0.98\textwidth, height=0.70\textheight]{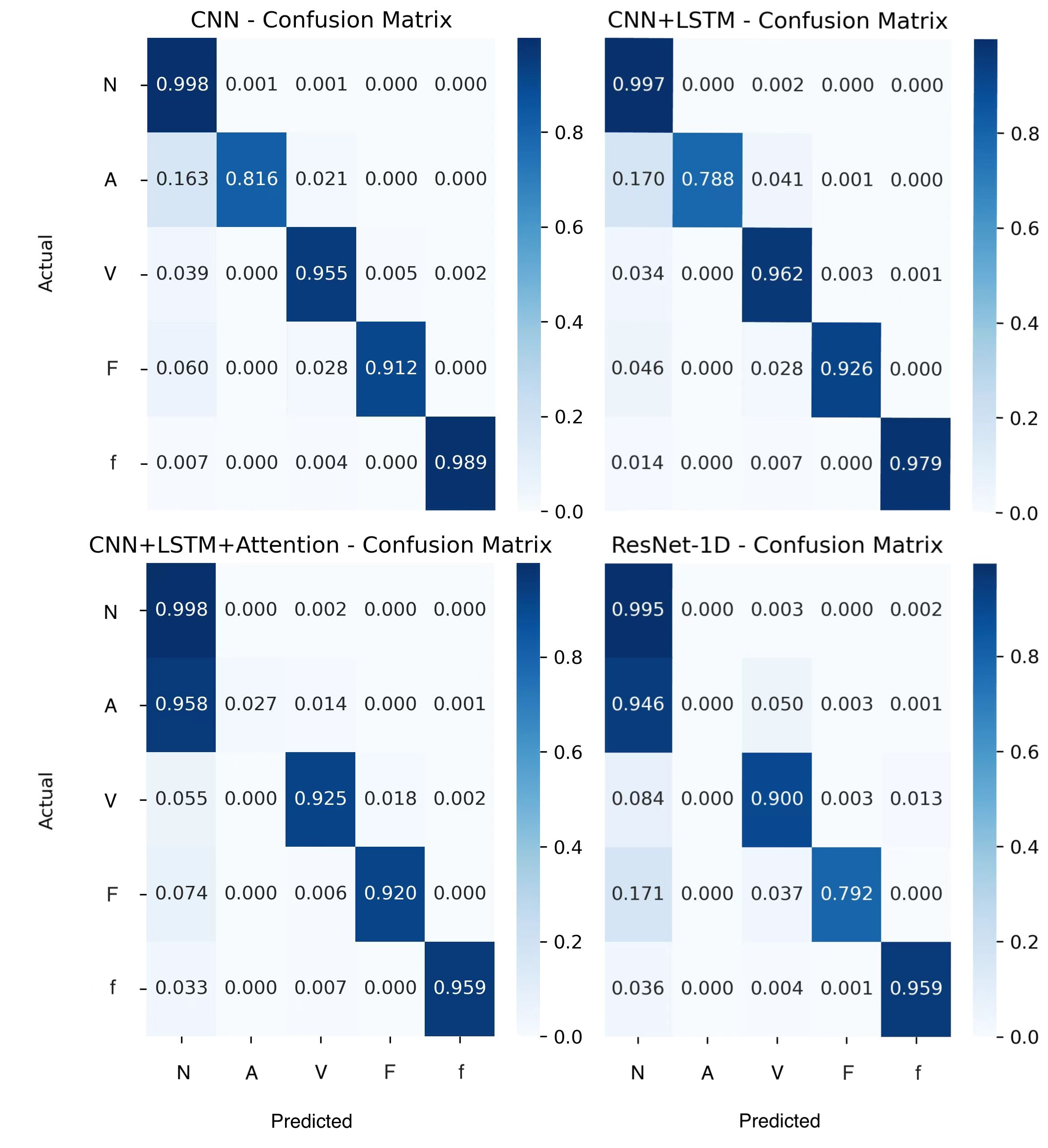}
    \caption{Confusion matrix of  CNN, CNN-LSTM, CNN-LSTM-Attention, and ResNet models on the test set. The labels represent: Normal (N), Atrial Premature (A), Premature Ventricular Contraction (V), Fusion of Ventricular and Normal (F), and Fusion of Paced and Normal (f).
}
    \label{fig:confusion_matrix_compare}
\end{figure}

Figure~\ref{fig:confusion_matrix_compare} presents the normalized confusion matrices of the four evaluated architectures (CNN, CNN-LSTM, CNN-LSTM-Attention and ResNet-1D) across the five ECG classes. Each matrix is row-normalized to highlight per-class recognition accuracy and the main misclassification patterns.

The CNN model demonstrates strong overall diagonal dominance, with accuracies exceeding 0.95 for most rhythm categories. However, approximately 16\% of samples from the second class were misclassified as the first, reflecting moderate overlap in waveform morphology and limited separability between these rhythm types.

The CNN-LSTM model achieves the most balanced and reliable performance among all architectures. Its diagonal accuracies remain above 0.92 for the major rhythm types, with values of 0.962 and 0.926 for the third and fourth categories, respectively, both of which exceed those of the CNN baseline. The confusion matrix exhibits a compact and sharply defined diagonal with minimal cross-class leakage, indicating that temporal recurrence substantially enhances both recall and overall stability.

In contrast, the CNN-LSTM-Attention model exhibits uneven results: while most classes exceed 0.90 accuracy, one minority class collapses to 0.027, indicating over-sensitivity of the attention mechanism under data imbalance.

The ResNet-1D model performs least effectively, with one category completely misclassified (0.000) and noticeably dispersed off-diagonal elements. Although residual connections improve optimization stability, the model fails to capture long-range temporal dependencies essential for ECG rhythm discrimination.

Overall, the CNN-LSTM architecture provides the best trade-off between accuracy, recall, and robustness. Its confusion matrix reflects consistent per-class recognition and minimal inter-class overlap, verifying that the combination of convolutional feature extraction and recurrent temporal modeling yields the most dependable performance across all ECG categories.

\subsubsection{ROC Curve Comparison}

\begin{figure}[htbp]
    \centering
    \includegraphics[width=\textwidth]{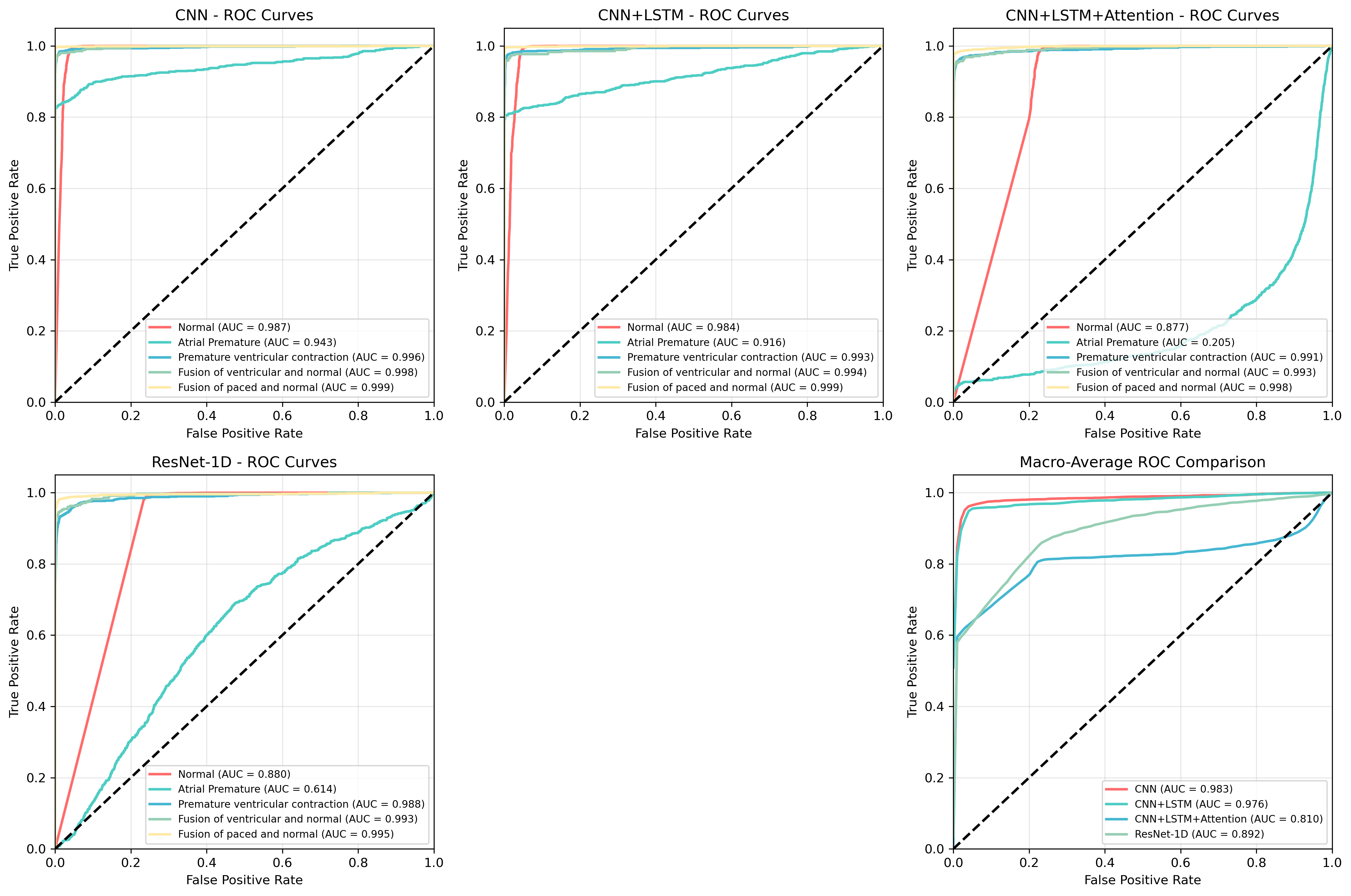}
    \caption{ROC curves and AUC comparison for all arrhythmia classes}
    \label{fig:roc_curves}
\end{figure}

To comprehensively evaluate the discriminative performance of different architectures, Receiver Operating Characteristic (ROC) curves and the corresponding Area Under the Curve (AUC) metrics (Figure~\ref{fig:roc_curves}) were analyzed for all classes and models. The ROC curves illustrate each model’s sensitivity–specificity trade-off and highlight its ability to correctly distinguish arrhythmia types under varying decision thresholds.

Across all models, Premature Ventricular Contraction (V), Fusion of Paced and Normal (f), and Fusion of Ventricular and Normal (F) achieved consistently high AUCs. f and F maintained values above 0.99 for every architecture, and V also performed strongly with AUCs ranging from 0.988 to 0.996, indicating clear separability and reliable detection for these rhythm types. The Normal (N) class showed high AUCs for the CNN and CNN-LSTM models (approximately 0.98), whereas the attention-based and residual architectures produced substantially lower values around 0.88. Atrial Premature (A) was the most challenging category: the CNN and CNN-LSTM models achieved moderately high AUCs (0.92-0.94), while the CNN-LSTM-Attention and ResNet-1D models experienced severe degradation, with AUCs falling to 0.20 and 0.61. This contrast reflects both the limited representation of A-type beats and their close morphological resemblance to normal rhythms.

The CNN model achieves a macro-average AUC of 0.983, with consistently high discrimination for V, f, and F (AUC \(> 0.99\)), slightly lower performance for N (AUC \(= 0.987\)), and a moderate reduction for the A class (AUC \(= 0.943\)). This suggests that while the CNN captures morphological differences effectively, its sensitivity to atrial premature activity remains comparatively limited.

The CNN-LSTM model demonstrates consistent performance across classes, with a macro-average AUC of 0.976. The inclusion of temporal recurrence allows the model to better capture inter-beat dependencies, supporting more reliable detection of abnormalities that depend on temporal context.

The CNN-LSTM-Attention model achieves AUCs above 0.99 for V, f, and F but shows reduced performance for N (AUC \(= 0.877\)) and severe degradation for A (AUC \(= 0.205\)). These differences indicate that the attention-based model provides strong discrimination for rhythm types with clear morphological patterns but is less stable on rhythms with subtle deviations from normal morphology, such as A.

The ResNet-1D model reaches a macro-average AUC of 0.892, performing well on morphologically distinctive classes but showing noticeably lower AUCs for minority arrhythmias such as A (0.614) and moderately reduced values for N (0.880). Its deeper residual structure supports hierarchical feature extraction but is associated with more variable class-wise performance, particularly in underrepresented categories.

Macro-average ROC analysis further supports these trends. The CNN and CNN-LSTM models achieve the strongest overall AUCs, reflecting more balanced and reliable discrimination across arrhythmia categories, while the CNN-LSTM-Attention and ResNet-1D models show substantially lower macro-level performance. These results reinforce the observation that architectures with stable convergence behavior also generalize more consistently across rhythm types.

Overall, the findings show that convolutional architectures capture stable ECG morphology, while the addition of LSTM layers significantly improves temporal discrimination and generalization. Attention mechanisms further enhance discrimination for rhythm types with well-defined waveform patterns but remain less stable for subtle arrhythmias. Macro-average ROC analysis confirms that the CNN-LSTM model offers the most balanced and robust performance among all evaluated architectures.

\subsubsection{Confidence Interval Comparison}

\begin{figure}[ht]
    \centering
    \includegraphics[width=0.85\textwidth]{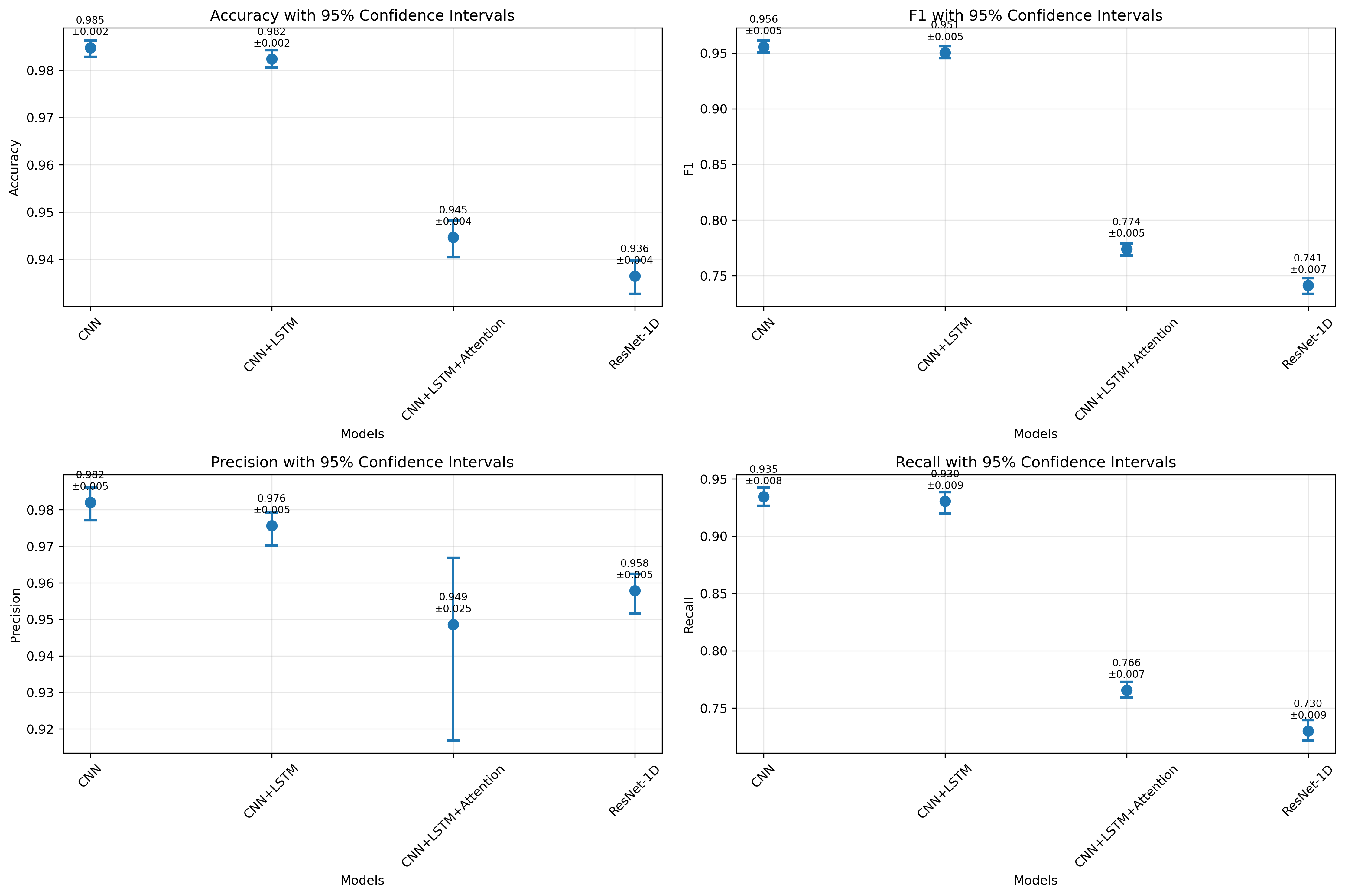}
    \caption{95\% confidence intervals for accuracy, precision, recall, and F1-score across models.}
    \label{fig:confidence_intervals}
\end{figure}

Figure~\ref{fig:confidence_intervals} illustrates the 95\% confidence intervals for the performance metrics of the four individual models, summarizing the mean and statistical variability across multiple experimental runs. Confidence intervals (CIs) provide insight into each model’s stability and robustness, highlighting not only their central performance trends but also the consistency of their learning behavior.

The CNN model exhibits narrow confidence intervals across all metrics, particularly for accuracy (0.985~\(\pm\)~0.002) and F1-score (0.956~\(\pm\)~0.005), indicating stable convergence and strong reproducibility. However, the slightly wider intervals in recall suggest sensitivity to class imbalance, as misclassification of minority arrhythmias introduces higher variance.

The CNN-LSTM model maintains comparably tight CIs, with accuracy around 0.982~\(\pm\)~0.002 and F1-score near 0.951~\(\pm\)~0.005, but achieves greater overall generalization consistency. The integration of recurrent layers effectively captures temporal dependencies, improving inter-beat discrimination and reducing cross-class variability compared with the CNN baseline.

The CNN-LSTM-Attention model achieves similar mean accuracy (approximately 0.945) but with broader confidence ranges, particularly in recall (0.766~\(\pm\)~0.007) and F1-score (0.774~\(\pm\)~0.005). This reflects its higher model complexity and sensitivity to initialization or data imbalance, especially in minority classes such as Atrial Premature. While attention enhances discriminative focus, it also amplifies variance when salient features are sparsely distributed.

The ResNet-1D model shows the largest confidence intervals among all architectures, with accuracy (approximately 0.936~\(\pm\)~0.004) and recall variability exceeding \(\pm 0.009\). This suggests that deeper residual layers may lead to overfitting or unstable optimization under limited data conditions, affecting reproducibility across random seeds or folds.

In summary, the CNN-LSTM architecture achieves the best trade-off between stability and generalization, maintaining tight confidence intervals while preserving temporal discrimination capability. Although CNN demonstrates slightly narrower intervals, its lack of sequential modeling limits adaptability. These results reinforce that CNN-LSTM delivers the most reliable and consistent performance among individual models, whereas more complex variants, such as Attention-based and ResNet-1D architectures, introduce greater variability.

\subsubsection{Grad-CAM Visualization}

\begin{figure}[ht]
    \centering
    
    \centerline{\scriptsize (a) Sample 1}\includegraphics[width=0.7\textwidth]{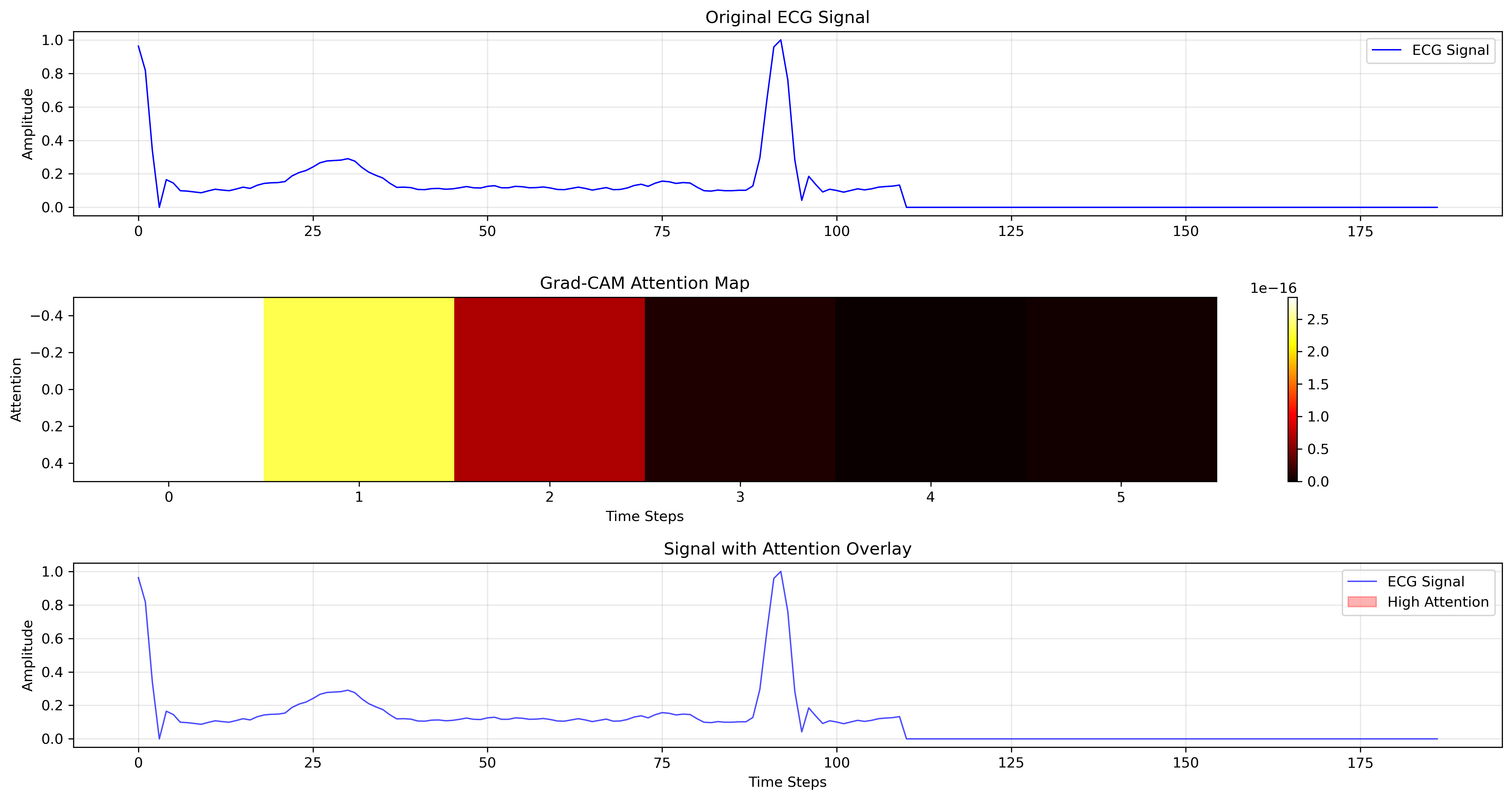}
    \vspace{0.5ex}

    \vspace{2ex}

    \centerline{\scriptsize (b) Sample 2}\includegraphics[width=0.7\textwidth]{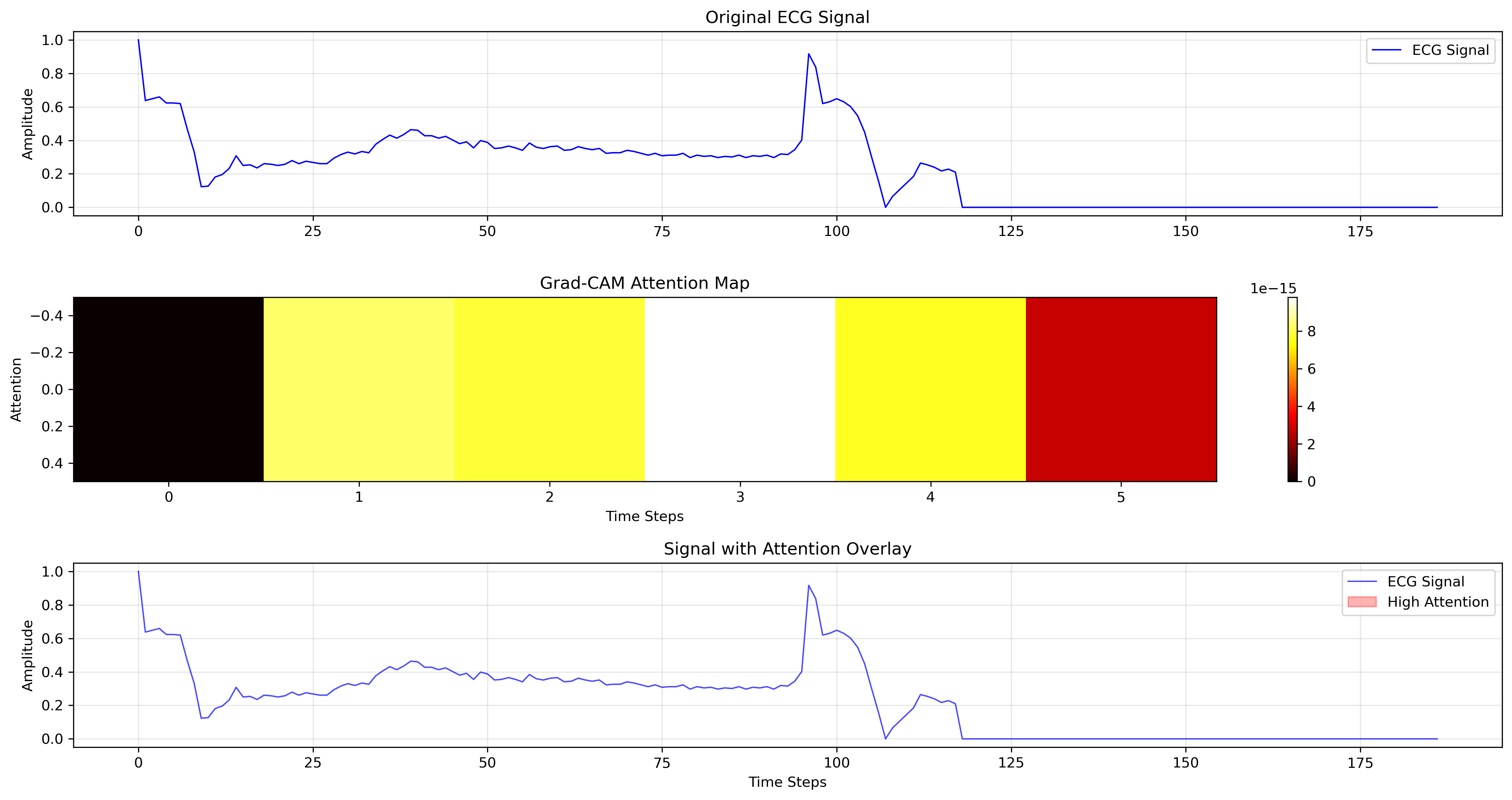}
    \vspace{0.5ex}

    \caption{Grad-CAM visualizations for representative normal and abnormal ECG beats.}
    \label{fig:gradcam_visualizations}
\end{figure}

Figure~\ref{fig:gradcam_visualizations} illustrates Grad-CAM \cite{selvaraju2017gradcam} visualizations for two representative ECG samples, revealing how the CNN-LSTM model attends to physiologically meaningful regions of the signal.

Sample 1 shows a strong concentration of attention around the QRS complex, particularly near the steep depolarization peaks, while the isoelectric baseline receives little focus. This indicates that the model primarily bases its classification on the depolarization phase, aligning with established diagnostic cues for normal cardiac rhythm.

Sample 2 exhibits a broader attention distribution, highlighting both the QRS complex and the subsequent ST-T segment. The extended high-attention regions correspond to repolarization-related morphology changes, such as altered ST slope or T-wave shape, suggesting that the model can recognize complex waveform patterns associated with abnormal beats.
Together, these results confirm that the CNN-LSTM architecture learns to focus on clinically relevant waveform components. The model shifts from highly localized attention in typical beats to more distributed attention in morphologically irregular signals, demonstrating interpretable and physiologically consistent reasoning.

\subsection{Impact of Ensemble Learning}

\begin{figure}[ht]
    \centering
    \includegraphics[width=0.85\textwidth]{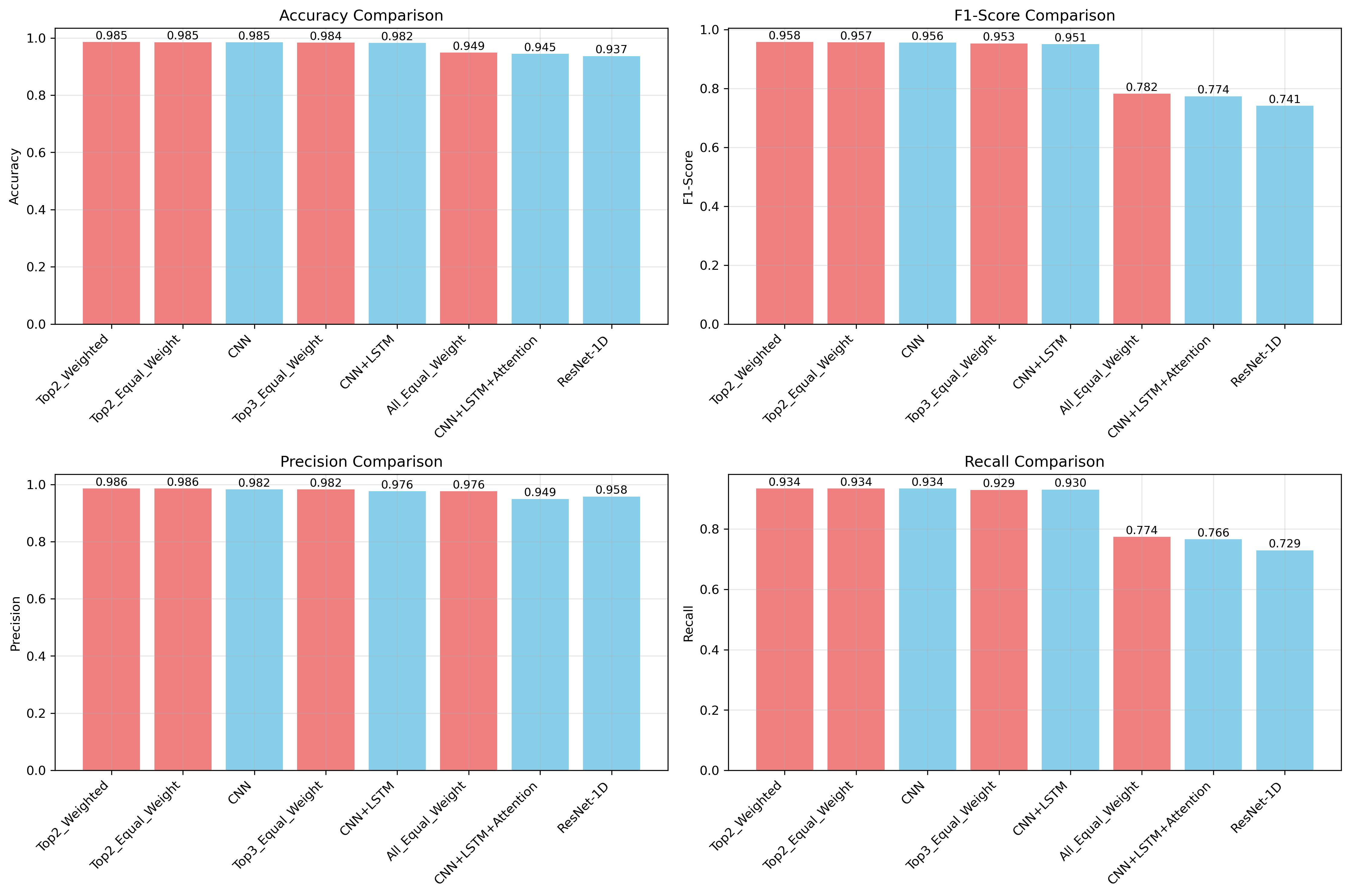}
    \caption{Ensemble model performance comparison across accuracy, precision, recall, and F1-score}
    \label{fig:ensemble_model_comparison}
\end{figure}

Based on the comparative results shown in Figure~\ref{fig:ensemble_model_comparison}, both individual and ensemble models were evaluated across four metrics: accuracy, precision, recall, and F1-score. While individual architectures such as CNN, CNN-LSTM, CNN-LSTM-Attention, and ResNet-1D demonstrate distinct strengths, their performance varies with class distribution and feature complexity. Among them, the CNN-LSTM model provides the most balanced overall performance, achieving high precision and recall across major rhythm categories, whereas the CNN model exhibits the most stable convergence and strong local feature extraction capability. Together, these two models represent the most complementary pair for ensemble integration.

The ensemble models further improve stability and class balance. The \textit{All Equal Weight} ensemble maintains strong accuracy and consistency, achieving approximately 0.949 in accuracy and 0.782 in F1-score, demonstrating that uniform aggregation effectively averages out the biases of individual networks. The \textit{Top3 Equal Weight} ensemble slightly enhances recall while maintaining similar precision, confirming that including three high-performing models captures more complementary patterns without overfitting. The \textit{Top2 Equal Weight} ensemble, combining the CNN and CNN-LSTM models, achieves a recall of 0.934 and demonstrates enhanced sensitivity to minority arrhythmias, accompanied by a slight decrease in precision due to its higher responsiveness to subtle signal variations.

The \textit{Top2 Weighted Ensemble} delivers the most balanced and robust performance among all fusion strategies. By assigning weights proportional to the validation F1-scores of the CNN and CNN-LSTM models, it achieves an overall F1-score of 0.958, with a precision of 0.986 and recall of 0.934, indicating a slight bias toward higher precision while maintaining strong generalization across classes. This indicates that performance-based weighting effectively reinforces the discriminative strength of the best model while retaining useful complementary information from the secondary one. The improvement is particularly evident in the detection of rare arrhythmias, where variance across models is reduced and decision boundaries become more stable.

In summary, ensemble fusion provides a significant performance gain over individual models by enhancing generalization, mitigating class imbalance, and improving consistency. The \textit{Top2-Weighted} ensemble (F1-score: 0.958) stands out as the best-performing configuration, achieving the most favorable trade-off between precision and recall and demonstrating the strongest robustness for ECG rhythm classification.

\section{Conclusion and Future Work}

This study presented a comprehensive evaluation of multiple deep neural network architectures for electrocardiogram (ECG) classification, including CNN, CNN-LSTM, CNN-LSTM-Attention, and ResNet-1D models, along with several ensemble strategies. Through extensive experiments and analyses across multiple performance metrics, the results demonstrate that temporal modeling and attention mechanisms significantly enhance the discriminative capability of ECG classifiers, while ensemble fusion further improves robustness and generalization.

Among the individual models, the CNN-LSTM network achieved the most balanced performance, combining the spatial feature extraction of convolutional layers with the temporal modeling ability of recurrent units. The CNN model also performed strongly, offering efficient local pattern recognition and stable convergence. These two architectures together provided complementary strengths that formed the basis for the best-performing ensemble configuration.

The ensemble models further reinforced stability and class balance. In particular, the Top2 Weighted ensemble, composed of CNN and CNN-LSTM, achieved the best overall results with an F1-score of 0.958, with precision of 0.986 and recall of 0.934. By assigning weights proportional to the validation F1-scores of the two leading models, this strategy effectively leveraged their complementary representations and mitigated overfitting to class imbalance. The ensemble also improved recognition of minority arrhythmias and maintained consistent accuracy across all rhythm categories, confirming its effectiveness as a practical and scalable ECG classification framework.

In future work, several directions can be explored to further improve performance and interpretability. First, data augmentation and synthetic beat generation could be employed to address the scarcity of rare arrhythmia samples, improving model sensitivity to minority classes. Second, adaptive attention mechanisms may be developed to dynamically adjust focus across temporal regions, thereby enhancing the detection of subtle atrial activations. Third, lightweight and deployable architectures could be designed for real-time ECG monitoring applications, balancing accuracy and computational efficiency. Finally, integrating explainable AI techniques with clinical decision-support systems may facilitate trustworthy deployment in medical environments.

Overall, this work provides a systematic benchmark of deep learning architectures for ECG classification and demonstrates that combining temporal modeling, attention-based learning, and ensemble fusion leads to a robust, interpretable, and clinically reliable framework for automated arrhythmia detection.

\section*{CRediT Authorship Contribution Statement}
\textbf{Yun Song:} Conceptualization, Methodology, Software, Formal analysis, Writing - Original Draft, Writing - Review \& Editing, Project administration. 
\textbf{Wenjia Zheng:} Conceptualization, Methodology, Software, Formal analysis, Writing - Original Draft, Writing - Review \& Editing, Project administration. 
\textbf{Tiedan Chen:} Conceptualization, Investigation, Writing - Original Draft. 
\textbf{Ziyu Wang:} Data Curation, Visualization, Methodology, Formal analysis. 
\textbf{Jiazhao Shi:} Software, Validation, Resources. 
\textbf{Yisong Chen:} Data Curation, Investigation, Visualization.

\section*{Declaration of Competing Interest}
The authors declare that they have no known competing financial interests or personal relationships that could have appeared to influence the work reported in this paper.

\section*{Funding}
This research did not receive any specific grant from funding agencies in the public, commercial, or not-for-profit sectors.

\section*{Data Availability}
The dataset used in this study (MIT-BIH Arrhythmia Database) is publicly available at PhysioNet (https://doi.org/10.13026/C2F305).

\section*{Acknowledgements}
The authors would like to thank the creators of the MIT-BIH Arrhythmia Database for making the dataset publicly available.

\section*{Declaration of generative AI and AI-assisted technologies in the manuscript preparation process}
During the preparation of this work the author(s) used Google Gemini in order to improve the readability and language of the manuscript and assist with LaTeX formatting. After using this tool/service, the author(s) reviewed and edited the content as needed and take(s) full responsibility for the content of the published article.

\end{document}